\documentclass[11pt]{article}
\pdfoutput=1 
\usepackage{jheppub}
\usepackage[utf8]{inputenc}

\usepackage{graphicx}
\usepackage{color, graphics}
\usepackage{amsmath,amssymb,amsfonts}
\usepackage{verbatim}   
\usepackage{subfigure}  
\usepackage{acronym}
\usepackage{hyperref}
\usepackage{mathrsfs}
\usepackage{multirow}
 \usepackage{cancel}
 \usepackage{url}
 \usepackage{hyperref}
 \usepackage{units}

\newcommand{\MeV}{{\, {\rm MeV}}}
\newcommand{\GeV}{{\, {\rm GeV}}}
\newcommand{\TeV}{{\, {\rm TeV}}}
\newcommand{\PeV}{{\, {\rm PeV}}}

\let\OLDthebibliography\thebibliography
\renewcommand\thebibliography[1]{\OLDthebibliography{#1}
  \setlength{\itemsep}{3.5pt}}

\def\beq{\begin{equation}\begin{aligned}}
\def\eeq{\end{aligned}\end{equation}}

\begin{document}

\title{Symmetric and Asymmetric Reheating}
\author[a,b]{Edward Hardy,}
\emailAdd{ehardy@ictp.it}
\affiliation[a]{Abdus Salam International Centre for Theoretical Physics, Strada Costiera 11, 34151, Trieste, Italy}
\affiliation[b]{Department of Mathematical Sciences, University of Liverpool,Peach Street, Liverpool, L69 7ZL, U.K.}
\author[c]{James Unwin}
\emailAdd{unwin@uic.edu}
\affiliation[c]{Department of Physics,  University of Illinois at Chicago, Chicago, IL 60607, USA}

\abstract{
We study models in which the inflaton is coupled to two otherwise decoupled sectors, and the effect of preheating and related processes on their energy densities during the evolution of the universe. Over most of parameter space, preheating is not disrupted by the presence of extra sectors,  and even comparatively weakly coupled sectors can get an order 1 fraction of the total energy at this time. If two sectors are both preheated, the high number densities could also lead to inflaton mediated thermalisation. If only one sector is preheated, Bose enhancement of the late time inflaton decays may cause significant deviations from the perturbative prediction for their relative reheat temperatures.  Meanwhile, in Non-Oscillatory inflation models resonant  effects can result in exponentially large final temperature differences between sectors that have similar couplings to the inflaton.  Asymmetric reheating is potentially relevant for a range of beyond the Standard Model physics scenarios. We show that in dark matter freeze-in models, hidden sector temperatures a factor of $10$ below that of the visible sector are typically needed for the relic abundance to be set solely by freeze-in dynamics.}

\maketitle


\vspace{-2mm}

\section{Introduction}

Dark sectors decoupled from the Standard Model are well motivated both from UV and IR considerations. String compactifications that contain the Standard Model (SM) often also include dark sectors. Meanwhile, dark sectors can include viable dark matter (DM) candidates, with the possibility of self interactions \cite{Carlson:1992fn,Spergel:1999mh}, or distinct indirect detection signals \cite{ArkaniHamed:2008qn,Pospelov:2008jd,MarchRussell:2008tu}. However if a hidden sector contains light states, observable cosmology could be disrupted. For example, energy in light dark sector degrees of freedom increases the effective number of neutrino species $N_{\rm eff}$, and the current bound  on contributions beyond the SM is $\Delta N_{\rm eff}\lesssim 0.25~$ \cite{Cyburt:2015mya,Ade:2015xua}. New sub-keV states are further constrained by measurements of the Lyman-$\alpha$ spectrum, which gives the leading lower mass limit on thermal DM, $m_{\rm DM}\gtrsim3$ keV  \cite{Baur:2015jsy}. A straightforward way to relax tension with these observations is through the dark sector having a temperature $T_{\rm DM}$ substantially lower than that of the visible sector $T_{\rm SM}$  \cite{Hodges:1993yb,Kolb:1985bf}.

Further, some DM models, as well as other motivated beyond the SM  scenarios, need significant temperature differences between the visible sector and a dark sector. In freeze-in DM the dark and visible sectors are almost decoupled \cite{Hall:2009bx}, and the DM relic density is produced by a small interaction between the dark sector and the SM thermal bath. As we show, for the relic abundance to be determined only by the form of this interaction a dark sector that is initially significantly colder than the visible sector is required, even if the DM has annihilations to light hidden sector states. In particular, we impose that any annihilations are frozen out before the freeze-in starts. Models in which the DM is a glueball of a strongly  coupled hidden sector gauge theory also need the visible sector to be hot compared to the other sectors \cite{Carlson:1992fn}. Stable glueballs from such sectors are expected in some compactifications of string theory \cite{Faraggi:2000pv,Halverson:2016nfq,Harling:2007jy,Harling:2008px}. Additionally, differences between the energy in separate sectors at early times could have important consequences, independent of their final relative temperatures. For example, a sector having a high temperature in the early universe can lead to processes such as phase transitions or production of heavy or weakly coupled states \cite{Chung:1998zb,Greene:1997ge,Giudice:1999fb,Giudice:2001ep}.

The simplest reheating mechanism is perturbative inflaton decays, the speed of which is fixed by the single inflaton particle decay rate \cite{Guth:1980zm,Abbott:1982hn,Dolgov:1982th}. If the inflaton dominantly decays to a particular sector, this will be reheated to a higher temperature than others \cite{Berezhiani:1995am}, with a resulting temperature ratio that has the approximate parametric dependence $T_{1}/T_{2} \sim \sqrt{g_{1}/g_{2}}$, where $g_i$ is the inflaton coupling to kinematically accessible states in the two sectors.\footnote{Large differences in inflaton couplings could come from, for example, models with sequestering in an extra-dimension \cite{Barnaby:2004gg,Harling:2007jy,Harling:2008px}.} However, in models with relatively large couplings between the inflaton and other states, non-perturbative processes can lead to resonantly enhanced energy transfer during a period prior to perturbative decays, called preheating.  The dynamics of this has been studied extensively in single sector models  \cite{Traschen:1990sw,Shtanov:1994ce,Kofman:1994rk,Kofman:1997yn,Giudice:1999fb}. 
   
In this paper, we study the impact of preheating, and other associated processes, on the relative energy in two sectors that are decoupled, apart from both being coupled to the inflaton. The effects are complex, involving non-perturbative dynamics and particle distributions that are far from thermal equilibrium. To make progress we have to resort to a range of approximations and for preheating itself numerical simulations are useful. Consequently we do not obtain precise results, aiming instead to identify interesting physical processes that can be important. We also do not attempt to study complete realistic models of the visible and hidden sectors. Typically we just take two copies of a simple toy model sector, identical other than their coupling to the inflaton, and try to find their relative energy densities as a function of their couplings to the inflaton as the universe evolves. Already this leads to several possible effects, although further study of more complex models would be interesting.

One issue is whether the dynamics of preheating to a sector are modified by the presence of other sectors coupled to the inflaton also being preheated. This can happen since backreaction or rescattering from a second sector might cause preheating to end sooner than it otherwise would. Immediately after preheating, thermalisation between sectors by off-shell inflaton scattering could also change the distribution of energy, potentially enhanced by the high number and energy densities produced by preheating. In many models there will subsequently be a long period of matter domination, due to the energy remaining in inflaton states. In this case, the late time relative temperature of two sectors will be close to the perturbative prediction unless inflaton mediated thermalisation is efficient, although as noted the early time dynamics might still lead to physically important effects.\footnote{Late time inflaton mediated thermalisation is important in many models, and a careful analysis of this can be found in \cite{Adshead:2016xxj}.} 
However, if the inflaton has large trilinear couplings it might decay perturbatively before matter domination is reached, and the dynamics of this could be affected by preheating. In such models, several complicated effects might occur, and final temperatures either closer together or further apart than perturbative inflaton decays alone are possible. As an example we study the effect of Bose enhancement of the inflaton decay rate. 
 
We also note that exponentially large differences in the final temperatures of sectors with order 1 differences in their couplings to the inflaton are possible in Non-Oscillatory models of inflation \cite{Felder:1998vq,Felder:1999pv}. In such theories, there are no perturbative inflaton decays. Meanwhile the energy transferred non-perturbatively is exponentially suppressed below a threshold coupling, and large temperature differences can occur without tuning parameters to particular values.
 
The structure of the paper is as follows: In Section \ref{sec:2} we motivate studying temperature difference between sectors by calculating the relative initial temperature ratio needed for the DM relic abundance to be set solely by freeze-in. In Section \ref{sec:3} we analytically and numerically study preheating in models with the inflaton coupled to two sectors. In Section \ref{sec:4} we consider effects such as thermalisation that can take place once preheating is finished. In Section \ref{sec:6} we describe how Non-Oscillatory inflation models with instant preheating can lead to large temperature ratios between sectors. Finally we discuss phenomenological implications of our work in Section \ref{sec:7}.


\section{Initial conditions for freeze-in dark matter}
\label{sec:2}

In models of freeze-in dark matter \cite{Hall:2009bx,Cheung:2010gj,Chu:2011be,Yaguna:2011qn,McDonald:2001vt,Blennow:2013jba,Chu:2013jja,Elahi:2014fsa}, prior to freeze-in the abundance of DM is assumed negligible and the DM is generated by a portal operator connecting the hidden and visible sectors.  The final relic abundance is controlled only by the DM mass and the portal operator if this is renormalisable (otherwise production is UV sensitive and may depend on the SM reheat temperature).

Unless the inflaton has no decays to the dark sector, reheating leads to a population of DM, which we call the primordial component. For freeze-in to set the DM relic density, the primordial component of DM  (PDM) should be substantially less than the DM abundance observed today
\beq
\left.\frac{\Omega_{\rm PDM}}{\Omega_{B}}\right|_{\rm today} \ll
\left.\frac{\Omega_{\rm DM}}{\Omega_{B}}\right|_{\rm observed} \simeq5~.
\label{cond0}
\eeq
The conditions after reheating needed to satisfy this depend on the details of the hidden sector. We consider two example scenarios, depending on if the dark sector has number changing interactions.

\subsection{Dark sectors with no number changing interactions}
\label{2a}

One possibility is that reheating occurs via perturbative decays and the DM has no number changing interactions other than those involving the inflaton or freeze-in. If each inflaton decay produces order one DM state, then after reheating 
\beq
n_{\rm DM}^{\rm (RH)} 
\simeq {\rm Br}_{\phi\rightarrow{\rm DM}}~ n_{\phi}\Big|_{H=\Gamma_\phi}
\simeq {\rm Br}_{\phi\rightarrow{\rm DM}}\frac{\rho_\phi}{m_\phi}\Big|_{H=\Gamma_\phi}~,
\label{2.2}
\eeq
where ${\rm Br}_{\phi\rightarrow{\rm DM}}=\Gamma_{\phi\rightarrow{\rm DM}}/\Gamma$ is the inflaton ($\phi$) branching fraction to DM, $m_{\phi}$ is the inflaton mass, and $\rho_\phi$ is the inflaton energy density at $H\simeq\Gamma$, with  $H$ the Hubble parameter.

Assuming that the visible sector dominates the entropy of the universe (and that there are no further entropy injections, for example through a late decaying scalar) the initial condition for freeze-in DM, Eq.~\eqref{cond0}, is then
\beq
\begin{aligned}
{\rm Br}_{\phi\rightarrow{\rm DM}}
&\ll 5\times 10^{-10} \left(\frac{10^{10}~{\rm GeV}}{T_{\rm SM}^{\rm (RH)}}\right)
\left(\frac{10^{3}~{\rm GeV}}{m_{\rm DM}}\right)
\left(\frac{m_\phi}{10^{13}~{\rm GeV}}\right)
~,
\label{con1}
\end{aligned}
\eeq
where $T_{\rm SM}^{\rm (RH)}$ is the visible sector reheat temperature. For simplicity, we have assumed that the visible sector consists of just the SM at all energy scales up to $T_{\rm SM}^{\rm (RH)}$. Since the baryon asymmetry is small ($({n}_B - {n}_{\bar{B}})/s \simeq  0.88 \times 10^{-10}$ at the present time) a large temperature asymmetry is needed unless the reheat temperature is very low. Immediately after being produced by perturbative inflaton decays, the DM states will typically have extremely high kinetic energy. If the dark sector has non-number changing interactions, this could be redistributed amongst the DM including the freeze-in component. The warming generated by this is negligible, provided the branching ratio is $\ll 1$ as is required by the relic density constraint almost all of parameter space. Otherwise if there are no interactions at all, the primordial DM will either red-shift to be non-relativistic or form a small hot DM component, depending on the parameters of the model.

\subsection{Dark sectors with annihilations}
\label{2b}

Alternatively the DM might have number changing interactions, and we focus on the case in which it thermalises with a bath of light dark sector states. The contribution of the relic energy density of the new light states to the effective number of neutrinos can be sufficiently small, provided the hidden sector is slightly colder than the visible sector (which is needed for freeze-in anyway) \cite{Zentner:2001zr,Hasenkamp:2012ii}. Alternatively, it might be possible to construct models such that they decay to the SM. If the dark sector states are relatively heavy $\gtrsim 100~\MeV$, constraints from collider observations easily allow decays before BBN \cite{Jedamzik:2006xz}. Meanwhile, for lighter masses collider and astrophysics bounds require much longer decay times, and there is greater danger of observable energy injection \cite{2013PhRvD}, although again this may be allowed if the hidden sector temperature is relatively low. However, further study is required to ensure the coupling between the sectors does not disrupt freeze-in, and we leave this for future work. 

While the dark sector is at sufficiently high temperatures, number changing interactions keep the DM number density close to its equilibrium value. However, as the hidden sector temperature drops, the DM abundance will freeze-out, similarly to the usual WIMP scenario. In order that the DM relic abundance is set purely by the freeze-in operator, there are two constraints that must be satisfied. First, the component of DM left from thermal freeze-out must be substantially smaller than the observed quantity. Second, DM annihilations must be negligible by the time freeze-in of the DM relic density happens (and remain small at later times).\footnote{A slightly weaker possible constraint is that annihilations of the thermal population must finish before freeze-in, allowing annihilations after the freeze-in component is generated. Such models are interesting \cite{Cheung:2010gj,Chu:2011be}, and the constraints obtained on the hidden sector initial temperature are close to those we obtain.} Assuming no entropy injections or large changes in the effective number of relativistic degrees of freedom before freeze-in, the ratio of hidden sector and visible sector temperatures is a constant $\xi = T_{\rm DM}^{\rm (RH)}/T_{\rm SM}^{\rm (RH)}$, determined by dynamics at an early time such as reheating. The fraction of the visible sector's energy transferred to the hidden sector per Hubble time through the freeze-in operator is largest when freeze-in is taking place, and at earlier times interactions through this coupling do not  increase the relative temperature of the hidden sector significantly.

The calculation of the hidden sector freeze-out from the Boltzmann equations is straightforward, and closely follows \cite{Scherrer:1985zt}. The underlying physics can be understood from an approximate analytic solution accurate to the \% level. Freeze-out finishes when the hidden sector temperature is
\beq \label{eq:fi1}
\frac{m_{\rm DM}}{T_{\rm DM}^{\rm (FO)}} =  \ln\left( \alpha \right)-\left(n+\frac{1}{2}\right) \ln\ln\left(\alpha\right) ~.
\eeq
The parameter $\alpha$ is
\beq
\alpha = 0.038 \left(n+1\right) \frac{g}{\sqrt{g_{\rm v}}} m_{\rm DM} M_{\rm Pl} \left(\sigma v\right)_0 \xi^2~,
\eeq 
where $M_{\rm Pl}$ is the (unreduced) Planck mass, and $g_{\rm v}$ is the visible sector effective number of relativistic degrees of freedom, which are assumed to dominate the energy of the universe, and $g$ is the effective DM number of degrees of freedom (for bosons $g=g_{\rm DM}$, and for fermions $g=\frac{7}{8} g_{\rm DM}$). The annihilation cross section has been parameterised as $\langle \sigma v \rangle\equiv \left(\sigma v\right)_0 \left(m_{\rm DM}/T_{\rm DM}\right)^{-n}$. The primordial relic density gives a yield $y= n/s$ after freeze-out of
\beq \label{eq:fi2}
y_{\infty} = 3.79 \left(n+1\right) \frac{g_{\rm v}^{3/2}}{M_{\rm Pl} m_{\rm DM} \left(\sigma v\right)_0} \xi \left(\frac{m_{\rm DM}}{T_{\rm DM}^{\rm (FO)}}\right)^{\left(1+n\right)}~.
\eeq
From Eq.~\eqref{eq:fi1}, the freeze-out temperature is related to the DM mass by a parameter that is typically less than 1 and is a slowly varying function of the properties of the model, similarly to the usual WIMP case for which $T^{\rm FO}\simeq m_{\rm DM}/25$. 

\begin{figure*}[t!]
\includegraphics[width=0.495\textwidth]{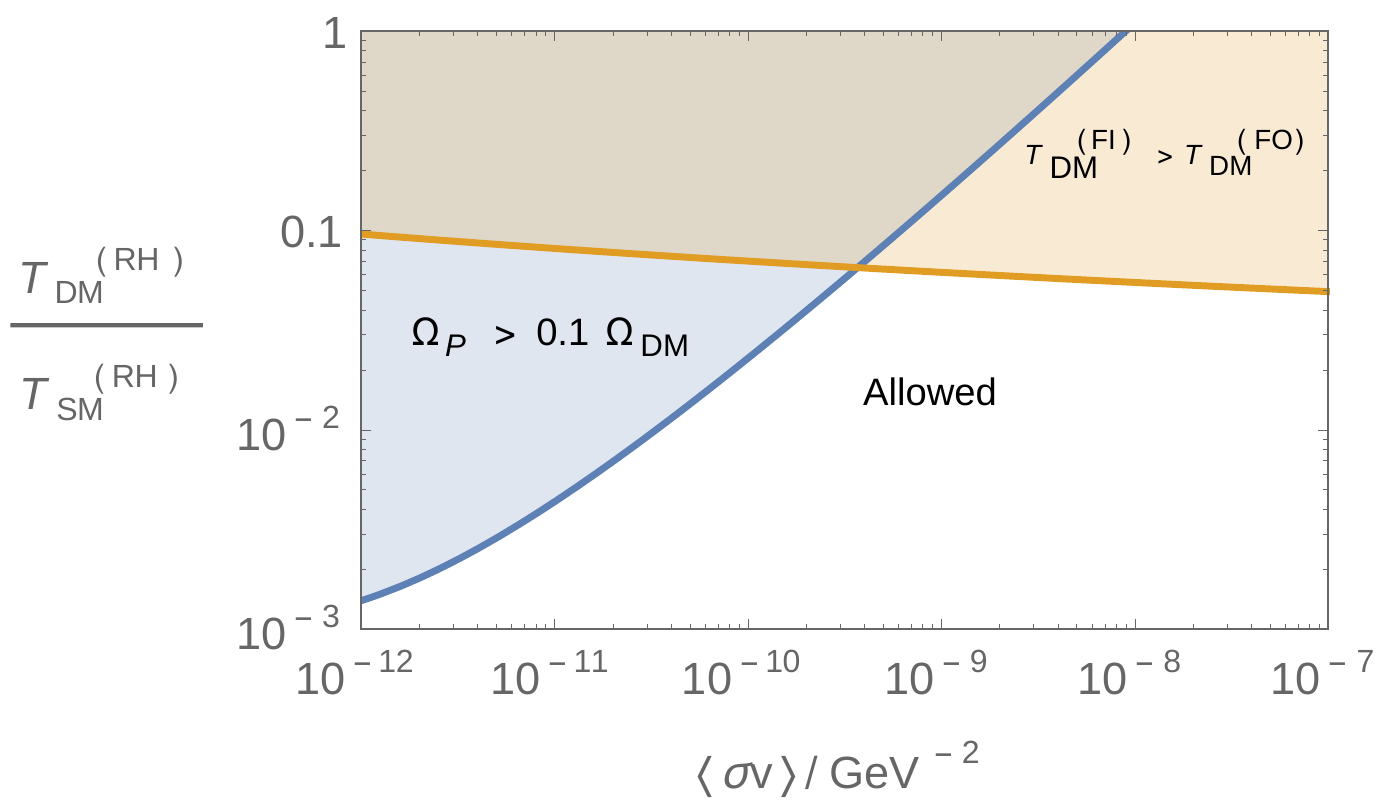}
\includegraphics[width=0.495\textwidth]{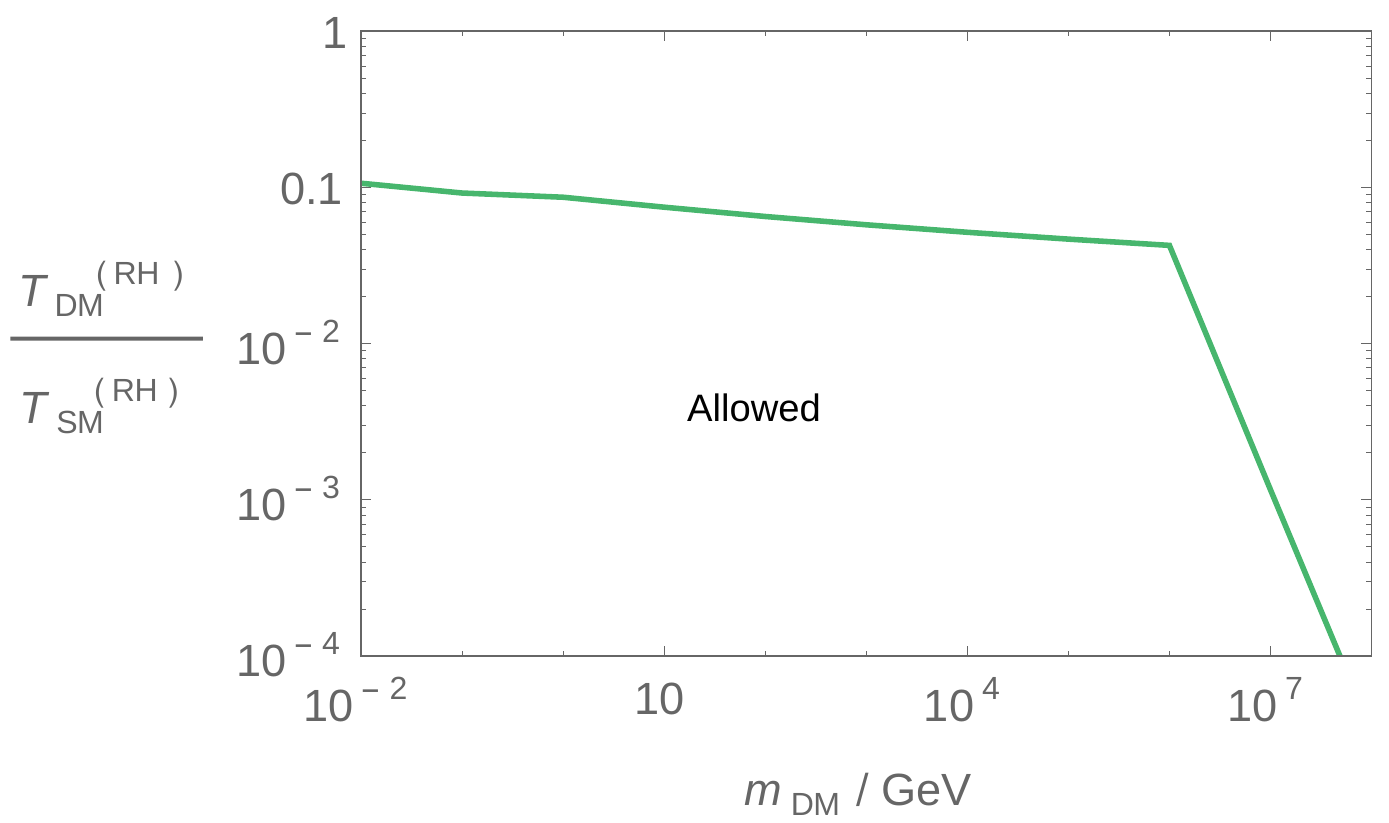}
\caption{{\bf \emph{Left:}} An example showing the interaction between the DM freeze-out cross section assuming s-wave annihilation and the parameter $T_{\rm DM}^{\rm RH}/T_{\rm SM}^{\rm RH}$, for a model with $m_{\rm DM} = 10~\GeV$. In the blue shaded region, the primordial relic density exceeds $0.1 \Omega_{\rm DM}$, while in the orange shaded region the DM annihilations freeze-out after freeze-in has occurred. In both cases the DM relic abundance is not set purely by the freeze-in mechanism. {\bf \emph{Right:}} The maximum allowed value of $T_{\rm DM}^{\rm RH}/T_{\rm SM}^{\rm RH}$ as a function of the DM mass, if the relic abundance is to be set by freeze-in alone, assuming s-wave annihilation. At high DM masses $\gtrsim 100~\TeV$ the unitarity bound on the annihilation rate results in strong constraints on the hidden sector temperature.
\label{fig:2}}
\vspace{2mm}
\end{figure*}

The relic density and freeze-out temperature constraints together lead to a maximum ratio of hidden sector and visible sector temperatures $\xi$, apart from for very large DM masses, for which the unitarity bound on annihilations is stronger \cite{Griest:1989wd}. The largest $\xi$ are allowed if freeze-in takes place as late as possible, corresponding to models in which freeze-in happens at a temperature $T_{\rm SM} \simeq m_{\rm DM}$. In this case, the temperature constraint and Eq.~\eqref{eq:fi1} requires that $\xi < T_{\rm DM}^{\rm (FO)}/m_{\rm DM}$ (which is an implicit equation since the freeze-out temperature depends on $\xi$). The relic density constraint is imposed by demanding that $\Omega_{\rm PDM}< 0.1~ \Omega_{\rm DM}$ in Eq.~\eqref{eq:fi2}. In Fig.~\ref{fig:2} (left) we illustrate the interaction of these constraints, as a function of the annihilation cross section, assuming $n=0$ and a DM mass of $10~\GeV$. 

The maximum allowed $\xi$  as a function of the DM mass is plotted in Fig.~\ref{fig:2} (right), for s-wave annihilation. At small masses this is approximately constant because  of the logarithmic dependence of Eq.~\eqref{eq:fi1} on the parameters of the model, and for DM masses below a PeV, the hidden sector must be colder by  a factor $\sim 10$ relative to the visible sector. For masses above $\sim \PeV$  the unitarity bound, $\left<\sigma v_{r}\right> < 4\pi/\left(m_{\rm DM}^2 v_{r}\right)$  for s-wave annihilation and relative velocity $v_{r}$ \cite{Griest:1989wd}, constrains the maximum possible cross-section, and much smaller hidden sector temperatures are needed. The unitarity bound becomes important at slightly higher masses than the usual WIMP case because, for a given DM mass, the constraint on the annihilation freeze-out requires a smaller annihilation cross section than that which would lead to the correct DM relic abundance for $\xi=1$. Light DM masses $\lesssim 100 ~\MeV$, and hidden sector 
states for them to annihilate into, may require colder hidden sector temperatures to avoid constraints in complete models. 

In other models, freeze-in will dominantly occur at temperatures $T_{\rm SM}^{\rm (FI)} > m_{\rm DM}$, leading to stronger constraints on $\xi$. For example freeze-in through a portal generated by a heavy state, with mass above the DM mass, that leads to a non-renormalisable operator linking the two sectors when integrated out.
 In this case, the bound on the freeze-out temperature is to order of magnitude given by
\beq
\begin{aligned}
\xi &\lesssim \frac{T_{\rm DM}^{\rm (FO)}}{T_{\rm SM}^{\rm (FI)}}  \sim \frac{m_{\rm DM}}{20~ T_{\rm SM}^{\rm (FI)}}~,
\end{aligned}
\eeq
since the hidden sector freeze-out temperature is typically $T_{\rm DM}^{\rm (FO)} \sim m_{\rm DM}/20$. Precise bounds can be straightforwardly calculated. Meanwhile, if freeze-in is via non-renormalisable operators that are not UV completed below the reheating temperatures, the dominant freeze-in production occurs immediately after reheating. Consequently, for such models number changing interactions must be absent at all relevant temperatures and the constraints are as for Section \ref{2a}.

If a hidden sector thermalises through $2\rightarrow n$ and $n \rightarrow 2$ self interactions, rather than with a bath of light states, the dynamics are changed. In such models the DM number density decreases over time, resulting in the hidden sector heating up. For the motivated case of the glueball of a hidden sector being the DM, the relic density constraint is 
\beq
\frac{T_{\rm DM}^{\rm (RH)}}{T_{\rm SM}^{\rm (RH)}} \sim 10^{-2} \left(\frac{\MeV}{\Lambda}\right)^{1/3}~,
\eeq
where $\Lambda$ is the hidden sector confinement scale \cite{Halverson:2016nfq}. To avoid observational constraints on DM self-interactions $\Lambda \gtrsim 100 ~\MeV$ is needed \cite{Boddy:2014yra}, requiring a large initial temperature asymmetry.


\section{Preheating in models with two sectors}
\label{sec:3}

In many models with relatively large couplings between the inflaton and other states, the first stage of reheating  happens through a period of parametric resonance \cite{Kofman:1994rk}.\footnote{There is a danger that large couplings to other fields cause radiative corrections to the inflaton potential, making it harder to realise inflationary models with preheating, however we do not worry about this issue.} In this section we apply existing results from single sector models to  preheating in theories with multiple inflaton decay sectors. An especially interesting aspect is whether there are regions of parameter space where efficient preheating would be expected in the case of only one sector, but there is no efficient energy transfer because of the dynamics from the other sector.

We focus on models in the broad resonance regime (defined shortly), since in an expanding universe the alternative of narrow resonance does not transfer a significant fraction of the inflaton energy. As is well known, preheating  terminates early when an order 1 fraction of the total energy in still in inflaton modes, and this remains true in models with multiple sectors. The system then evolves in a complicated way, with the expectation that each sector will move towards a thermal distribution. There is also the possibility of extra energy transfer from the inflaton during this time, and narrow resonances could be important for the dynamics. However, due to the high number density of decay products, which have energy distributions far from equilibrium, at this stage the system must  be studied using lattice simulations. Ultimately  for reheating to complete perturbative inflaton decays, and therefore trilinear inflaton couplings, are needed \cite{Dolgov:1982th,Dufaux:2006ee}. These decays will have an important effect on the final relative temperatures of the two sectors and we study this, and other late time processes, in Section \ref{sec:4}.

In broad resonance, the coherent oscillations of the inflaton act as a changing background mass $m_{\chi}$ for the inflaton's decay products $\chi$ (the non-expectation value induced mass of $\chi$ must be smaller for efficient resonance). Over most of an oscillation, the evolution of the decay products' mass is slow compared to $1/m_{\chi}$, and close to adiabatic, so that particle production is weak. However, when the inflaton is near  the minimum of its potential, $m_{\chi}$ is small and evolving fast. Consequently the system can be highly non-adiabatic, and the number density of the decay products may grow extremely fast. Assuming the decay products are bosons also results in a Bose enhancement and exponential growth in number density. In particular, during preheating, one oscillation of the inflaton leads to particle production 
\beq \label{eq:np1}
n_{k}^{i+1} \simeq e^{2\pi \mu_{k}^i} n_{k}^i ~,
\eeq
where $n_{k}^{i}$ is the number density of the momentum mode $k$ of $\chi$ after $i$ inflaton oscillations, and $\mu_k^i$ is a numerical coefficient. $\mu_k^i$ itself is a fast fluctuating function of the momentum $k$, the velocity of the inflaton in the non-adiabatic part of its potential, and the coupling between the inflaton and $\chi$ \cite{Kofman:1997yn}. Preheating to fermionic decay products is also possible, due to similar non-adiabatic processes. However in this case, though the rate of particle production is parametrically faster than perturbative decays, the exponential growth in number density is absent \cite{Giudice:1999fb,Greene:2000ew}.

If the rate of resonant energy transfer exceeds that of perturbative reheating, and happens sufficiently fast that the produced modes can reach large occupation number before being redshifted away from the resonant momentum band, preheating will have a significant effect.  The details of resonance in an expanding universe are complex because the decrease in the inflaton oscillation amplitude from the expansion of the universe can lead to the parameter $\mu_k$ in Eq.~\eqref{eq:np1} changing significantly even during the period of a single inflaton oscillation. However, since particle production only take place during a small fraction of each oscillation, analytic results are possible, and these have been studied carefully in \cite{Kofman:1997yn,Greene:1997fu} (for subsequent literature see the reviews \cite{Bassett:2005xm,Allahverdi:2010xz,Amin:2014eta}, and references therein).

As an example that can be analysed analytically, and is also compatible with current observations, we study a simple model with inflaton $\phi$ and potential
\beq
V= \frac{1}{2} m_{\phi0}^2 \phi^2 +  \frac{1}{2}  g_1^2 \phi^2 \chi_1^2 + \frac{1}{2}  g_2^2 \phi^2 \chi_2^2~,
\label{eq:pot}
\eeq
where  $\chi_1$ is a real scalar in one sector, and $\chi_2$ a real scalar in a second otherwise decoupled sector. A quadratic inflaton potential is commonly found in, for example, Chaotic Inflation models  \cite{Linde:1983gd}, and preheating in the single sector version of this model has been carefully studied in \cite{Kofman:1997yn} where full details may be found. Rather than attempt to review all details, we simply quote the relevant physical results. The inflaton mass is fixed to $m_{\phi 0}=10^{-6} M_{\rm Pl}$, motivated by the COBE normalisation for a pure quadratic inflaton potential \cite{Bunn:1996py}, and the masses of $\chi_{1,2}$ are assumed to both be much less than $m_{\phi 0}$. For simplicity, we take the same initial conditions as \cite{Kofman:1997yn}  $t_0=\pi/\left(2 m_{\phi 0}\right)$ and $\Phi\left(t_0\right)= 2 M_{\rm Pl}/\sqrt{3 \pi^3}$.  While the number density of the produced states is sufficiently small that they do not significantly affect the inflaton equation of motion, 
$\Phi\left(t\right) =  \Phi\left(t_0\right) t_0/t $ due to the expansion of the universe, where $\Phi\left(t\right)$ is the magnitude of the inflaton oscillations.

\subsection{Single sector preheating}

Properties of the resonances are determined by a parameter
\beq \label{eq:qt}
q_i(t)= \frac{g_i^2 ~\Phi\left(t\right)^2}{4m_{\phi}\left(t\right)^2}~,
\eeq
where $\Phi\left(t\right)$, and $m_{\phi}\left(t\right)$ includes contributions from expectation values, which decreases with time. Since we are considering models with two different decay products, these will have separate $q_i$ determining the resonances to that sector. If $q_i\gg 1$,  resonances are wide, with width $\Delta k \sim m_{\phi}$, and in an expanding universe $\mu_k$ fluctuates randomly within one inflaton oscillation due to $\Phi\left(t\right)$ changing. This regime is known as stochastic resonance, and the number density of a mode is determined by the effective average value of a random parameter $\mu_k\left(t\right)$. Meanwhile for smaller values of $q_i$, but still $\gtrsim 1$, and at later times when the universe is expanding slower, the values of the $\mu_k$ fluctuate less. The resonance bands remain wide, but the dynamics are closer to broad resonance in a non-expanding universe with the $\mu_k$ constant and calculable over timescales longer than $1/m_{\phi}$. In both the stochastic and broad resonance regimes particle creation dominantly occurs at times when $\phi \left(t\right) \simeq 0$ and the evolution is non-adiabatic. For later use we also note that the typical momentum of modes produced is 
\beq
k_{*,i} \simeq \sqrt{\frac{1}{2} g_i \Phi\left(t\right) m_{\phi}}~.
\eeq

In contrast, if $q_i \lesssim 1$ resonance bands are narrow, with width parametrically $\Delta k \sim q_i m$, and $\mu_k \sim q_i/2$ for the resonance with smallest $k$. However, for small $q\left(t\right)$ energy is not transferred efficiently, since momentum modes are redshifted out of the resonance band faster than they are generated, preventing an exponential growth. In particular, this happens if $q_i^2 m_{\phi} > H$  \cite{Kofman:1997yn}, and for the model we consider $H \sim m_{\phi}/10$, so fast energy transfer does not occur for $q\lesssim 1/4$.\footnote{The condition that resonance is faster than perturbative decays is satisfied if $q m \gtrsim \Gamma$ where $\Gamma$ is the perturbative decay rate, and is typically less constraining.} Since $q\left(t\right)$ decreases, a theory starting with $q\left(t_0\right) \gg 1$ will begin in the stochastic resonance regime, and then probably pass to broad resonances (unless $q\simeq$ 1 is quickly reached, while the universe is still expanding fast). This transition is not important for our purposes since its effects can be absorbed in the averaged $\mu_k$ values. Subsequently, the system will evolve into the narrow resonance regime, and shortly after exponentially efficient energy transfer will end, so evolution to a narrow resonance regime is a crucial event. 

If the dominant effect reducing $q_i\left(t\right)$ is Hubble expansion decreasing $\Phi\left(t\right)$, the end of preheating happens at a time 
\beq \label{eq:tH}
t_{H,i} = \frac{g_i \Phi \left(t_0\right)}{m_{\phi0}} t_0 ~.
\eeq
Alternatively, the decay products of the inflaton can themselves modify the properties of the resonance and potentially stop it. One important effect, called backreaction, is the generation of an expectation value $\left<\chi_i^2\right>$, which affects the inflaton equation of motion. A careful analysis of the impact on the inflaton oscillations is needed since $\left<\chi_i^2\right>$ is itself time varying, and leads to  $\left<\chi_i^2\right> \simeq n_{\chi i}/\left(g_i \Phi  \right)$ (where $n_{\chi i}$ is the total number density of $\chi_i$) and
\beq \label{eq:phivevm}
\Delta m_{\phi}^2 \simeq \frac{g_i n_{\chi_i}\left(t\right)}{\Phi\left(t\right)}~.
\eeq
Importantly for our work, backreaction can potentially lead to two distinct stages of preheating (both of which take place in the stochastic/broad regime). 

Considering a model with only a single inflaton decay product, and $q\left(t_0\right) > 1$, first there will be a period when the contribution to the inflaton mass Eq.~\eqref{eq:phivevm} is negligible. If $q\left(t_0\right)$ is relatively small but still $\gtrsim 1$, this will be the only stage and $q\sim 1/4$ will be reached before the number density $n_{\chi}$ is large enough for the expectation value induced inflaton mass to be comparable to $m_{\phi 0}$. 

For larger $q\left(t_0\right)$, backreaction will be important and at a time $t_{b}$ leads to $\Delta m_{\phi}^2 \simeq m_{\phi 0}^2$. After this, the effective inflaton mass squared will increase exponentially fast, due to the exponential increase in $n_{\chi}$. Therefore the inflaton will oscillate much faster, and $q\left(t_f\right) \simeq 1/4$ will be reached in $\lesssim 10$ oscillations of the inflaton for typical values of $q_{1}\left(t_{b}\right)$, which take place over a very short time $t_f-t_b \ll t_b-t_0$. If $t_f$ is reached, the total kinetic energy in $\chi$ states is $\sim k_*^2 \left<\chi^2\right>$. Using Eq.~\eqref{eq:qt}, $m_{\phi} = g \Phi\left(t_f\right)$ and $\left<\chi^2\right> \simeq \Phi\left(t_f\right)^2$, so at this time the energy is automatically shared equally between the inflaton potential, kinetic energy of the $\chi$ (and also the interaction energy $\sim g^2 \Phi^2 \left<\chi^2\right>$) \cite{Kofman:1997yn}. This is unlike if $q =1/4$ is reached before backreaction, in which case $m_{\phi}$ is independent of $\left<\chi^2\right>$, and the energy in $\chi$ states is much smaller. 

Preheating could also be stopped by scattering effects where a large $\left<\phi^2\right>$ develops due to the production of finite momentum $\phi$ states from interactions with the $\chi_i$. This would increase the mass of $\chi$, and for models with large couplings could stop resonance before $t_b$ (we consider this possibility later). Depletion of the inflaton zero mode by scattering will also reduce the energy transfer rate, however typically the resonance is stopped by backreaction before this becomes important.

\subsection{Two sector preheating and backreaction}
\label{sub3.2}

Having reviewed this known physics, we now consider models with two sectors, and in particular if backreaction leads to effects between sectors. For simplicity we assume that $g_1 \gtrsim g_2$ and the equivalent averaged parameters for the fastest growing modes (that is, the ones that determine the eventual $\chi_i$ number density) for each sector satisfy $\mu_1 \gtrsim \mu_2$, where a subscript labels the sector, not momentum. The extension to other cases is straightforward. For couplings $g_i$ in the range $10^{-4} \div 10^{-3}$ typical values of $\mu$ are distributed randomly in the range $0.1 \div 0.14$, with fluctuations up to $\mu \sim 0.2$ for some narrow windows of couplings. We also assume throughout that $g_i \gtrsim 10^{-6}$ so that $q_i\left(t_0\right) \gtrsim 1$ and at least some preheating can happen, otherwise the corresponding sector simply has no energy transferred during preheating.

The simplest scenario is if both the first and the second sector reach $q_i =1/4$ and finish preheating before backreaction is reached. This approximately happens if
\beq
g_i \mu_i \lesssim 3\times 10^{-4} \times 0.13~,
\eeq
for both sectors. In this case, the energy transferred from the inflaton to the $\chi_i$ is small, and the presence of another sector has no effect. During this stage of preheating, the number densities of $\chi_i$ are given by \cite{Kofman:1997yn}
\beq
n_{\chi i}(t)\sim10^{-4}\left(\frac{g_i^3 M_{\rm Pl}^3}{m_{\phi}^2\mu_i t^5}\right)^{\nicefrac{1}{2}}
e^{2\mu_i m_{\phi} t}~,
\label{eq:nchi2}
\eeq
and the time for $q_i = 1/4$ to be reached for each sector is given by Eq.~\eqref{eq:tH}.
Therefore the post-preheating relative number densities of the $\chi_1$ and $\chi_2$ is
\beq
\frac{n_{\chi 1}\left(t_{H,1}\right)}{n_{\chi 2}\left(t_{H,1}\right)} \simeq \left(\frac{g_1}{g_2}\right) \left(\frac{\mu_{2}}{\mu_{1}}\right)^{1/2} e^{6\times10^5 \left(\mu_1 g_1 - \mu_2 g_2 \right)} ~,
\eeq
where the number density of $\chi_2$ has been evolved from $t_{H,2}$ to the later time $t_{H,1}$. Since this scenario assumes $ 3\times10^5 \mu_i g_i \lesssim 12$, the exponent can have a large, but not enormous, effect. The total kinetic energy in the state $\chi_i$, which is $\simeq  k_{*,i}^2 \left<\chi_i^2\right>$, can be compared to the remaining energy in the inflaton oscillations 
\beq \label{eq:kerel}
\frac{k_{*,i}^2 \left<\chi_i^2\right>}{\frac{1}{2} m_{\phi}^2 \Phi\left(t_{b,0}\right)^2} \simeq \left(\frac{g_i}{g_0}\right)^{5/3} \left(\frac{\mu_0}{\mu_{i}} \right)^{1/2} e^{-24 \left(1-\frac{\mu_i g_i}{\mu_0 g_0} \right)} ~,
\eeq
where parameters are normalised relative to $g_0= 3\times 10^{-4}$ and $\mu_0 = 0.13$. For comparison between models, both the kinetic energy density and the inflaton energy in  Eq.~\eqref{eq:kerel} are evaluated at a time $t_{b,0}$, although the qualitative features of the result are not sensitive to this choice.\footnote{The interaction energy between $\phi$ and $\chi$ will be larger than the kinetic energy by a factor $\sqrt{\Phi\left(t_1\right)/m_{\phi}}\sim 100$, but redshifts faster due to the faster decrease of $\Phi\left(t\right)$ (this can be seen in lattice simulations  
\cite{Figueroa:2016wxr}).}   Due to the early termination of the resonance, only a small fraction of the inflaton energy is transferred during preheating in this scenario, unless a sector is very close to this boundary of reaching backreaction.

The more interesting case is if a sector gets blocked by backreaction, which by assumption happens fastest for sector 1. We do not explicitly study the possibility that the dynamics in the two sectors are close enough to both contribute significantly to the blocking, but this is straightforward to include by taking Eq.~\eqref{eq:phivevm} to have two sources. 

If the second sector has a sufficiently small $q_2\left(t_0\right)$, resonance will stop before backreaction happens, in particular if $t_{H,2} < t_{b,1}$.   The time for sector 1 to reach backreaction $t_{b,1}$ can be well approximated by \cite{Kofman:1997yn}
\beq \label{eq:cond}
\begin{aligned}
t_{b,1} & \simeq \frac{1}{4\mu_1 m_{\phi0}} \ln \left(\frac{10^6}{g_1^5} \right)~,
\end{aligned}
\eeq
and $t_{H,2}$ is given by Eq.~\eqref{eq:tH}. $t_{b,1}$ is only weakly dependent on $g_1$, and $\mu_1$ typically does not vary by more than a factor $\sim 2$ at most. Therefore imposing $t_{H,2} < t_{b,1}$ is almost equivalent to requiring $t_{H,2} < t_{b,2}$, that is the second sector would not reach backreaction if it was the only sector.  In models in which this is satisfied, sector 1 gets an order 1 fraction of the energy density since it reaches $q_1 = 1/4$ in the backreaction regime. Meanwhile, the energy in sector 2 is given by Eq.~\eqref{eq:kerel}, and is exponentially suppressed for small couplings.\footnote{In the backreaction dominated stage, the inflaton oscillates faster, potentially allowing narrow resonances to be important. However, since $m_{\phi}\left(t\right)$ is increasing fast, the values of momentum that are on resonance are changing, and $q$ which is already small decreases further, so no significant energy will be transferred.}

If $g_1$ and $g_2$ are both relatively large $\gtrsim 3\times 10^{-4}$, then $q_2\left(t_{b,1}\right) \gtrsim 1/4$ and the second stage of preheating will transfer energy to both sectors. Energy transfer to sector 1 continues  until $m_{\phi}$ increases enough that $q_1=1/4$. Meanwhile the energy transfer to sector 2 will stop earlier once $q_2=1/4$. During this process $\Phi\left(t\right)$ decreases due to the energy transfer to the $\chi_i$, and can be calculated using conservation of energy (this stage lasts much less than a Hubble time). However, during this time $\Phi\left(t\right)$ will only drop by a factor of $1/q_1\left(t_{b,1}\right)^{1/4}$ which is $\sim$ few, and most of the decrease happens during the last one or two inflaton oscillations. Therefore, to get approximate analytic results we assume $\Phi$ has a constant value $= \Phi\left(t_{b,1}\right)$ during this period. 

The expectation value $\left<\chi_1^2\right>$  increases proportional to $e^{4 \pi \mu_1 N}$, where $N$ is the number of inflaton oscillations at this stage, and the $q_i$ decrease at the same rate. Consequently the total number of inflaton oscillations that each sector is preheated for during this stage $N_{1,2}$ are related by
\beq \label{eq:n1n2}
\frac{e^{4 \pi \mu_{1} N_1}}{e^{4 \pi \mu_{1} N_2}} \simeq \frac{g_1^2}{g_2^2}~.
\eeq
Therefore, using Eqs.~\eqref{eq:nchi2} and \eqref{eq:cond}, the final relative number density of $\chi_1$ and $\chi_2$ (at final time $t_f$) is approximately 
\beq
\begin{aligned} \label{eq:numratio}
\frac{n_{\chi 2}\left(t_f\right)}{n_{\chi 1}\left(t_f\right)} & \simeq \frac{n_{\chi 2}\left(t_{b,1}\right)}{n_{\chi 1}\left(t_{b,1}\right)} \frac{e^{4\pi\mu_{2} N_2}}{e^{4\pi\mu_{1} N_1}} \\
&\simeq \left(\frac{g_2}{g_1}\right)^{3/2+2\mu_2/\mu_1} \left(\frac{\mu_1}{\mu_2} \right)^{1/2} \left(\frac{10}{ g_1^{5/6}} \right)^{-3 \left(\frac{\mu_{1}}{\mu_2}-{1} \right)} e^{-4 \pi N_1 \left(\mu_1-\mu_2 \right)} ~.
\end{aligned}
\eeq
Provided $\mu_1$ is not too different to $\mu_2$ the exponentials are not enormous (since $N_1\lesssim 10$). Therefore, if an analytic analysis is accurate, in this part of parameter space there can be significant differences between the number density of states in each sector, but not far beyond those that would come from perturbative decays. The total kinetic energy density in each sector is approximately $\simeq k_{*,i} n_{\chi i}$. Since $k_{*,1}/k_{*,2} \simeq \left(g_1/g_2\right)^{1/2}$, and an order 1 fraction of the total energy finishes in $\chi_1$ kinetic energy, the approximate energy fraction in the $\chi_2$ sector is straightforwardly obtained from Eq.~\eqref{eq:numratio}.

To summarise, if preheating is assumed to be stopped only by backreaction, then a sector that would be preheated efficiently if it was the only one in a theory will typically still be partially preheated. However, if another sector is more strongly coupled to the inflaton, the energy transferred to the less strongly coupled sector will be suppressed by powers of coupling constants (and potentially exponentials of $\mathcal{O}(1)$ numbers). This is in contrast to if the model only contained the less strongly coupled sector, in which case it would get an order 1 fraction of the total energy.

\subsection{Rescattering and other effects}

However, as mentioned backreaction is not the only effect that can end preheating. Another possibility, called rescattering, is that scattering of $\chi$ states off the inflaton zero mode produces a significant density of finite momentum inflaton states. These can induce an effective mass for $\chi_i$, which if it is larger than the typical resonant momentum $k_{*,i}$, prevents efficient preheating to the $\chi_i$ sector. The computation of this process is subtle because only inflaton modes with momentum $k \gtrsim k_{*,i}/4$ (called hard modes) are distinguishable from the inflaton zero mode for the purposes of preventing the resonance. Again careful discussion may be found in \cite{Kofman:1997yn}. Briefly reviewing the relevant physics, preheating to $\chi_i$ is disrupted if the induced $\chi_i$ mass is sufficiently large
\beq
\begin{aligned} \label{eq:resc}
g_i^2 \left<\delta \phi^2\right>_{\rm hard} &\gtrsim \frac{1}{8}k_{*,i}^2 
\\&\gtrsim
 \frac{1}{16} g_i m_{\phi} \Phi\left(t\right)~.
\end{aligned}
\eeq
The hard inflaton modes have kinetic energy $\sim \frac{1}{2} k_{*,i}^2 \left<\delta \phi^2\right>$, and therefore the efficient initial stage of preheating stops when approximately $1/256$ of the total energy in the system is in this form. After this time energy transfer will continue through more complex dynamics at a slower rate \cite{Prokopec:1996rr,Berges:2008sr,Berges:2016nru,Micha:2004bv,Micha:2003ws}.
Following \cite{Kofman:1997yn}, since the energy of the hard modes comes from that of the $\chi_i$, a lower bound on the time at which this happens is the time $t_r$ at which $\simeq 1/256$ of the systems energy is in  $\chi_i$ kinetic energy. It can be straightforwardly shown that this corresponds to $\left<\chi^2\right> \simeq 1/256~ \Phi\left(t_r\right)^2$. From conservation of energy, at this time
\beq \label{eq:phitr}
\Phi\left(t_r\right) &\simeq 2.5 q_{1}\left(t_{b,1}\right)^{-1/4} \Phi\left(t_{b,1}\right) \\&
\simeq 2.5 \Phi\left(t_{f}\right)~,
\eeq
where $\Phi\left(t_{f}\right)$ is the amplitude the inflaton would have in the absence of rescattering if preheating finished at $t_f$ when $q_1 =1/4$ due to backreaction.

These estimates allow a simple, approximate, discussion of the effects of rescattering in a model with two sectors. During the final stage of preheating $\left<\chi_1^2\right>$ grows $\sim e^{4\pi \mu_1 N}$ where $N$ is the number of inflaton oscillations after $t_{b,1}$. If preheating continued until backreaction was reached, then $\left<\chi_1^2\right> \simeq \Phi\left(t_{f}\right)^2$. Therefore, using $\left<\chi^2\right> \simeq 1/256~ \Phi\left(t_r\right)^2$ and Eq.\eqref{eq:phitr}, rescattering will stop preheating to $\chi_1$ after approximately 
\beq
N_{r} \simeq N_{1}- \frac{1}{4 \pi \mu_1} \ln\left(10\right)
\eeq
second stage oscillations of the inflaton, where $N_1$ is the number of second stage oscillations that would occur in the absence of rescattering. Using Eq.~\eqref{eq:n1n2}, this means that is $g_1/g_2 \lesssim 3$,  preheating to $\chi_2$ will still be taking place when rescattering blocks preheating to $\chi_1$. Note however, the actual dynamics will be complicated and potentially significantly different to this expression. 

Rescattering can therefore have interesting effects on the relative energy densities of the two sectors. If $g_2$ is not much smaller than $g_1$, then $\chi_2$ will continue to be preheated after resonance to $\chi_1$ is stopped, because the mass induced by rescattering is $\sim g_i$ whereas $k_{*,i} \sim \sqrt{g_i}$. It is plausible that in a complete system the total energy transferred to $\chi_2$ could be comparable to that transferred to $\chi_1$ in such a case, despite the smaller coupling. However, our approximate understanding is far from this level of precision. Even if $g_2$ is substantially smaller than $g_1$, if $g_1 \gtrsim 0.01$ preheating will be ended by rescattering significantly earlier than backreaction would have stopped it \cite{Kofman:1997yn}. At this time most of the energy will be in the form of inflaton-$\chi_{i}$ interaction energy, and the  final ratio of energy transferred to each sector depends on subsequent complicated dynamics. Further, rescattering will first block resonances with momentum $\simeq k_{*}$, and efficient energy transfer could continue to higher momentum resonances.  In Section \ref{sub3.5}, we use a lattice simulation to study processes at later times.

If rescattering is effective, then at the end of preheating, any preheated states $\chi_i$ are likely to have a large mass $\sim k_{*,i}\left(t_r\right)$. This will change due to later interactions of the inflaton, and would also redshift away, but could be important for subsequent dynamics. Using expressions for $\Phi\left(t_r\right)$  from \cite{Kofman:1997yn}, immediately after preheating ends $m_{\chi_i}\sim \sqrt{g_i}~\times 10^{15}~\GeV$.

For reheating to complete, the energy remaining in the inflaton must be transferred to the sectors through perturbative decays, otherwise the universe will be matter dominated at late times. This requires trilinear couplings of the inflaton to $\chi_i$, in a full model. There are two especially motivated sizes of these couplings, given the quartic interactions of Eq.~\eqref{eq:pot}. If the inflaton potential is such that $\phi$ gets a VEV of the natural size $\left<\phi\right> \sim m_{\phi}$, the quartic interactions with $\chi$ will lead to trilinear interactions $V\supset g_i^2 m_{\phi} \phi \chi_i^2 $. Alternatively, if the underlying theory is supersymmetric, and the potential comes from a superpotential, then there will automatically be terms $\sim g_i m_{\phi} \phi \chi_i^2$ \cite{Dufaux:2006ee}.

In passing we note that if the theory comes from a superpotential $W = m_{\phi} \Phi^2+ g_1 \Phi X_1 X_1 + g_2  \Phi X_2 X_2 $ (where $\Phi$, $X_i$ are chiral superfields with $\phi$, $\chi_i$ their $\theta=0$ components), then there will automatically be a large coupling between $\chi_1$ and $\chi_2$ that is  $\sim g_1 g_2 \chi_1^2 \chi_2^2$. For all the values of couplings we consider, this will lead to fast thermalisation between the two sectors after preheating. Therefore in such models it is very hard to have a theory with separate sectors that both couple to the inflaton, but have different temperatures. Despite this, we continue to use the supersymmetry inspired relation between the quartic and trilinear couplings as a reasonable toy model for a theory with large trilinear couplings.

The presence of trilinear couplings can also affect preheating, or even be the dominant source of it. This has been studied analytically and numerically in \cite{Dufaux:2006ee}, and can lead to interesting difference to the quartic case, since the states $\chi_i$ can now get negative mass-squared parameters (preheating is also altered if couplings between the inflaton and its decay products have negative sign \cite{Greene:1997ge}). In the case of a trilinear with size $\sim g_i^2$, the quartic coupling is typically dominant throughout preheating. For a larger trilinear $\sim g_i$, the beginning of preheating is driven by the quartic interaction, but as the inflaton oscillation amplitude drops, the trilinear becomes increasingly important, eventually dominating the dynamics. Rather than attempt to analytically study a model containing multiple sectors coupled with large trilinear couplings to the inflaton, we instead simulate such a model numerically.

Finally, while we have focused on a simple model with well motivated parameter choices, it would also be interesting to study models with different inflaton masses. We expect the generic picture of backreaction and rescattering potentially leading to cross-sector effects to remains, but the details could be different. Other inflaton potentials might also lead to significant differences, though resonance is mostly sensitive to the dynamics of the inflaton around the minimum of its potential. More dramatic modifications are possible if the assumption that the two sectors are identical apart from the value of their coupling to the inflaton is relaxed, for example if there is a sector in which only fermions are coupled strongly to the inflaton. As mentioned, this can still undergo a form of preheating, and the dynamics of combining this with a preheated bosonic sector could be interesting.


\subsection{Numerical analysis}
\label{sub3.5}

Numerical simulations of preheating and the immediate aftermath can provide valuable information about a dynamical regime that cannot be easily studied analytically, for example on the energy transfer from the inflaton once backreaction and rescattering are important. Immediately after preheating the distribution of energy in both the inflaton and decay products sectors is dominantly in low momentum modes, compared to a thermal distribution with the same energy density. Scattering will move the system towards being thermalised, but the processes involved are complex   \cite{Salle:2000hd,Aarts:2000mg,Allahverdi:2000ss,Allahverdi:2002pu,Micha:2003ws,Micha:2004bv,Desroche:2005yt,Mukaida:2015ria}. Also a significant fraction of the total energy remains in the inflaton zero mode, and it is important for subsequent dynamics (discussed in Section \ref{sec:4}) if this remains the case, or if scattering moves it into high momentum inflaton modes.

Previously, there have been many developments in understanding of the relevant physics utilizing lattice simulations, for example \cite{Khlebnikov:1996mc,Khlebnikov:1996wr,Khlebnikov:1996zt,Felder:2000hr,Micha:2002ey,Podolsky:2005bw,Dufaux:2006ee,Felder:2006cc}.\footnote{In the case of a quartic inflaton potential conformal invariance allows the expansion of the universe to be removed by a field redefinition, and more analytic analysis of preheating and the subsequent thermalisation is possible \cite{Khlebnikov:1996mc,Greene:1997fu,Micha:2004bv}.} Further, recent progress has allowed for studies of dynamics at later times, and with finer grid spacing giving information on the behaviour of higher momentum states  \cite{Frolov:2008hy,Figueroa:2016wxr}. In order to determine the early stages of thermalisation, relatively large simulations are needed, and we do not carry out a full analysis here (in Section \ref{sec:4} we use  information from \cite{Podolsky:2005bw,Dufaux:2006ee}, assuming the results will be 
similar in models with multiple sectors).
Instead we implement models with multiple sectors in the code {\tt LATTICEEASY} \cite{Felder:2000hq}, to determine whether the analytic arguments based on the initial stages of preheating in Section \ref{sub3.2} hold once the full dynamics of the system are included. The simulation is carried out on a grid of spatial size $128^3$ (which is fairly small compared to the current best numerics). We run to a time $t=300/m_{\phi}$ by which point the fraction of energy in each sector is stable.

\begin{figure*}[t!]
\includegraphics[width=0.48\textwidth]{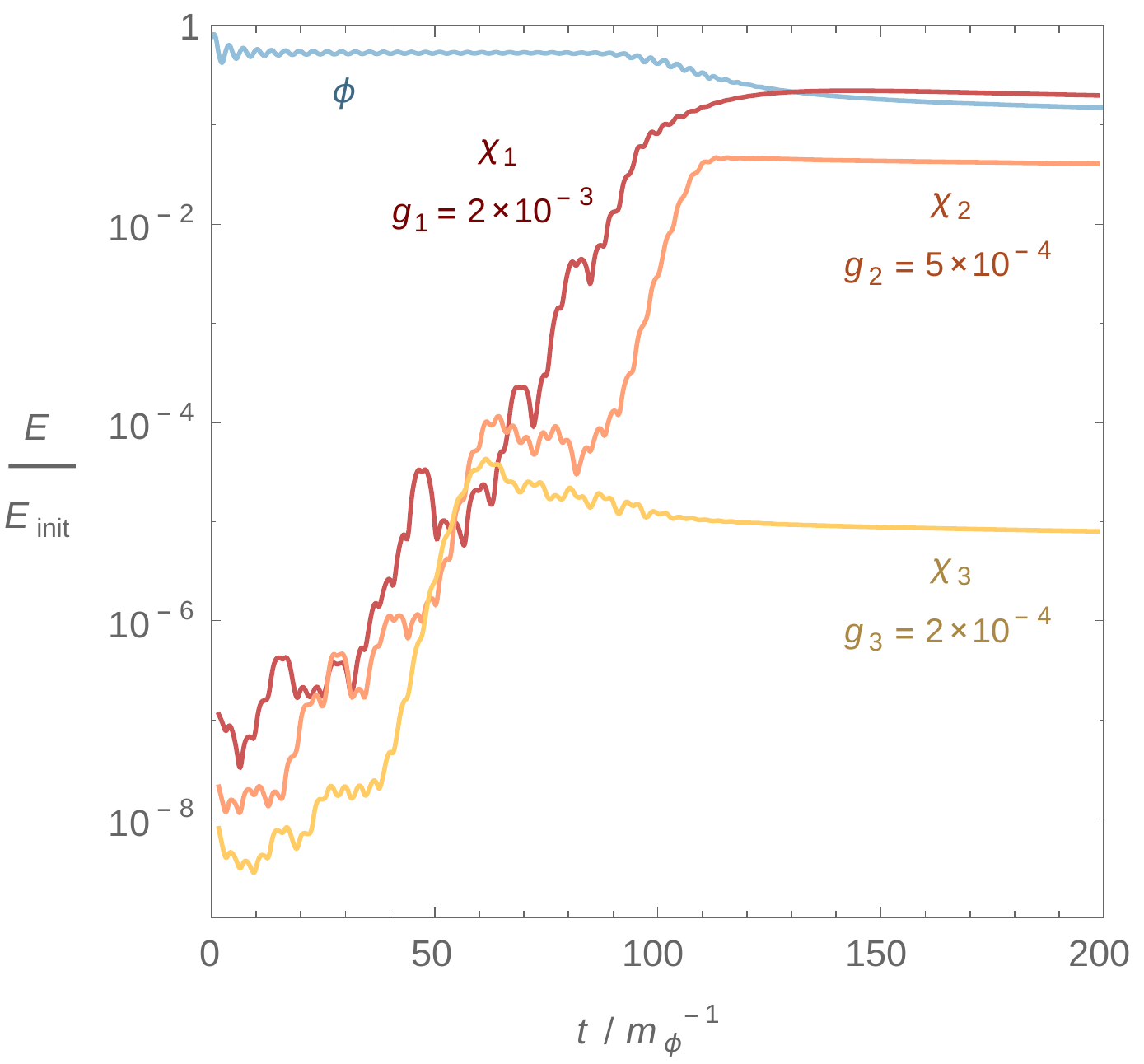}
\hspace{2mm}
\includegraphics[width=0.47\textwidth]{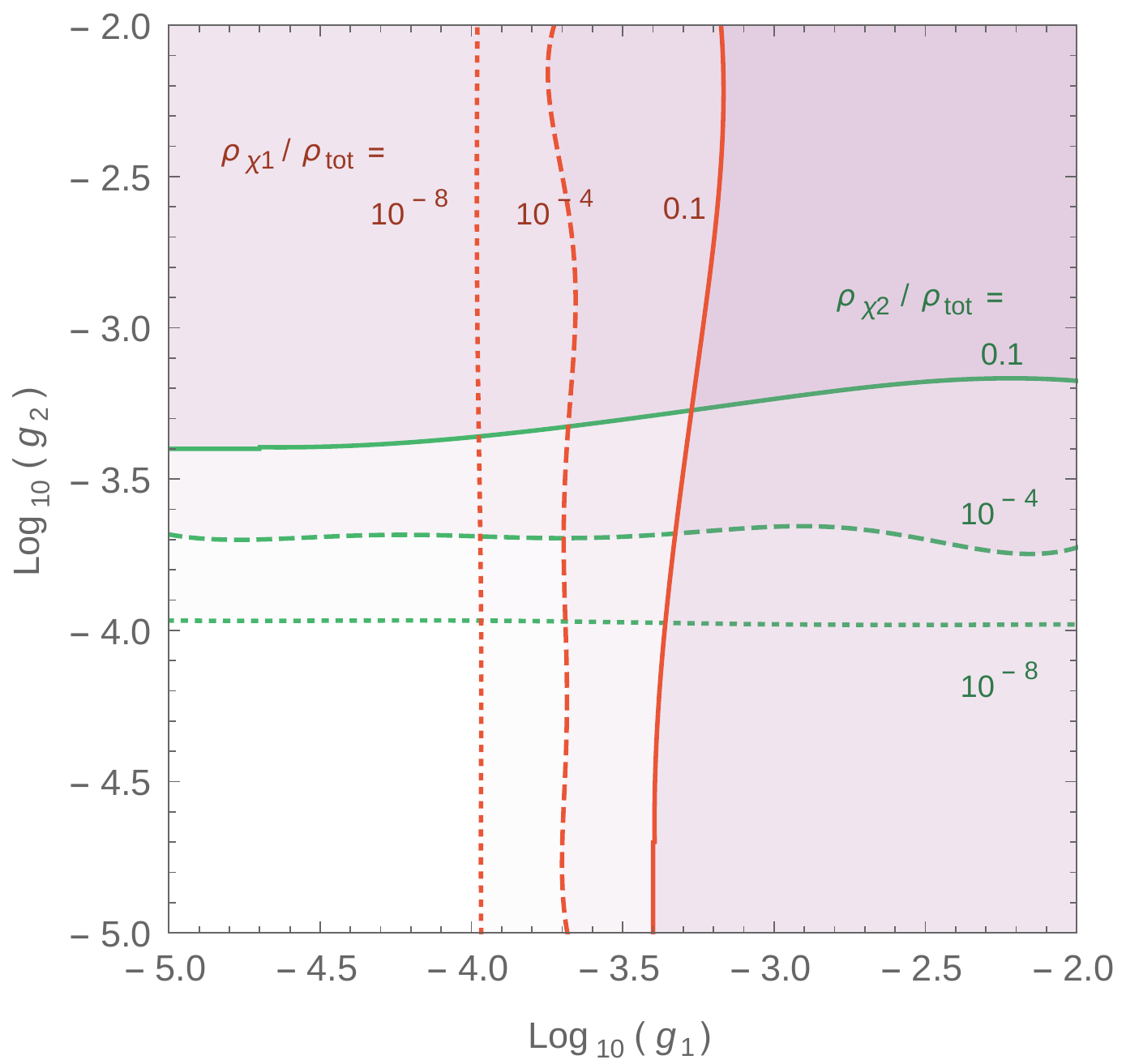}
\caption{{\bf \emph{Left: }} An example of the evolution of the energy (compared to the initial energy of the system $E_{\rm init}$) in each sector during preheating, for a model in which the inflaton can decay to three sectors. The inflaton potential is quadratic and couples to the other states by $\frac{1}{2} g_i^2 \phi^2 \chi_i^2$. Sector 1 is fully preheated, and despite its smaller coupling sector 2 also gets a significant fraction of the energy. The coupling to sector 3 is just below threshold for efficient preheating and the fraction of energy transferred is much smaller. For clarity, the plotted energy is averaged over 3 inflaton oscillations. {\bf \emph{Right:}} The fraction of a system's total energy transferred to each sector in a two sector model during preheating, as the inflaton couplings to the two sectors are varied. The inflaton potential is the same form as in the left panel. In the purple shaded region, one or both sectors is efficiently preheated. Since the contours are close to straight 
lines, the presence of another sector does not significantly block preheating, even if it is more strongly coupled to the inflaton.
\label{fig:ph}}
\end{figure*}

In Fig.~\ref{fig:ph} left panel, the model Eq.~\eqref{eq:pot} is extended with a third scalar $\chi_3$ coupled similarly to the inflaton. The total energy in each sector is plotted during preheating, which continues to times $\sim 100/m_{\phi}$, including the kinetic energy, gradient energy, and also the energy due to interactions with the inflaton for each $\chi_i$. The interaction energy is on slightly different footing, since it will decrease as the inflaton expectation value drops. However, the amount of  energy in this form is comparable to that in kinetic and gradient energies, so does not affect the results. At the start of preheating, transfer of energy happens to all three fields at a similar rate despite their different couplings, until a time $\sim 60/m_{\phi}$. At this point $\chi_3$, which has the smallest coupling, stops preheating, while the other two sectors continue. 

The coupling $g_2$ is such that $\chi_2$ would get approximately $1/10$ of the total energy during preheating if there were no other sectors present. This is still the case in the model including $\chi_1$ shown in  Fig.~\ref{fig:ph} left  (and remains true even for theories with larger values of $g_1$). At the end of the simulation, the difference in energy between $\chi_1$ and $\chi_2$ is slightly smaller than that which would come from perturbative 
reheating. For larger $g_2$, the energy transfer to $\chi_2$ is increased, again approximately independent of the value of $g_1$. Meanwhile the energy in $\chi_3$ is highly suppressed, which is also the case in models containing only $\chi_3$ given the value of $g_3$. The low proportion of the energy in $\chi_3$, despite its coupling to the inflaton not being much smaller than that of $\chi_2$, shows the sharp threshold between efficient and inefficient preheating.

In the right panel, we plot contours for which the states $\chi_1$ and $\chi_2$ get a fixed fraction of the total energy, in a two sector model. The contours are almost horizontal and vertical over the parameter range accessible to our simulation, indicating that the presence of a second sector has a limited effect on the preheating of the first, compatible with the left panel. There is a minor increase in the value of coupling needed to get an order 1 fraction of the total energy density when a second sector is efficiently preheated. However this is significantly smaller than that predicted in the previous section assuming preheating is terminated by backreaction. In the limit that the second sector is decoupled from the inflaton, the values of $q_{1}\left(t_0\right)$ that lead to order 1, and suppressed, energy transfer are compatible with the analytic arguments, and also recent careful numerical analysis with large simulations \cite{Podolsky:2005bw,Figueroa:2016wxr}. Although the plot is shown up to couplings of $g_i=10^{-2}$, close to this value numerical errors in the simulations become more problematic because typical resonance modes are higher momentum and further from the IR dynamics (more discussion can be found in \cite{Figueroa:2016wxr}). 

In the left panel, a significant fraction of the energy transfer from the inflaton to $\chi_2$ happens at relatively late times, once the more strongly coupled $\chi_1$ has almost finished preheating (at a time around $100/m_{\phi}$), and prior to this there is a short period where energy transfer halts. 
These features are common when a sector is relatively weakly coupled but still preheated, and can also appear in models containing a single decay product. A possible explanation for the smallness of the cross-sector effects is that a significant fraction of energy transfer to a fairly weakly coupled sector during preheating happens after rescattering has become important. Even though a more strongly coupled sector causes rescattering to happen earlier, this late time energy transfer might not be disrupted. As mentioned, rescattering will affect the more weakly coupled sector less than the more strongly coupled sector, and there is likely to be a significant proportion of energy remaining in the inflaton zero mode. Further investigation of the dynamics, for example by analysing the spectrum of the inflaton and its decay products,  would be interesting.

We also consider models with relatively large trilinear couplings, and a potential
\beq \label{eq:trilineaL}
V=  \frac{1}{2} m_{\phi}^2 \phi^2 +  \sum_{i=1}^{2}\left[\frac{g_i}{2} m_{\phi} \phi \psi_i^2 + \frac{g_i^2}{2} \phi^2 \psi_i^2 + \frac{g_i^2}{8}\psi_i^4\right] ~,
\eeq
where all fields are real scalars. The relation between the couplings follows the supersymmetry inspired pattern (but as discussed without a cross sector coupling). Repeating the numerical study we find that the cross-sector effects are very similar to those shown in Fig.~\ref{fig:ph} for the model with only quartic couplings. The presence of a second preheated sector slightly increases the coupling needed to get an order 1 fraction of the total energy, but not dramatically. We have also repeated the numerical simulation for a model with only trilinear inflaton $\chi_i$ couplings, that is a potential Eq.~\eqref{eq:trilineaL} but without the $g_i^2 \phi^2 \chi^2$ terms. In this case the results are again similar, even though the dynamics of preheating are different  \cite{Dufaux:2006ee}.


\section{Post-preheating evolution}
\label{sec:4}

After the efficient energy transfer from the inflaton finishes, the remaining inflaton quanta will decay perturbatively, and effects at these times are crucial for the final relative temperatures of the hidden and visible sectors. In this section we discuss potentially relevant processes and estimate the parts of parameter space where they are significant. We stress that due to large uncertainty in the dynamics of this period even in the case of a single sector model, some of our results are necessarily rough approximations. 

Aside from the dynamics we discuss, the relative temperatures of separate sectors could also be altered by changes in the number of relativistic degrees of freedom $g$ at relatively late times, though the resulting effects are $\Delta T/T \sim \left(\Delta g/g\right)^{1/3}$ and typically not very large \cite{Feng:2008mu}. Alternatively, a non-relativistic scalar could dominate the energy of the universe at a time after reheating, and dominantly decay to a particular sector. This is a plausible scenario in models with flat directions, for example due to supersymmetry.

\begin{figure*}[t!]
\includegraphics[width=0.46\textwidth]{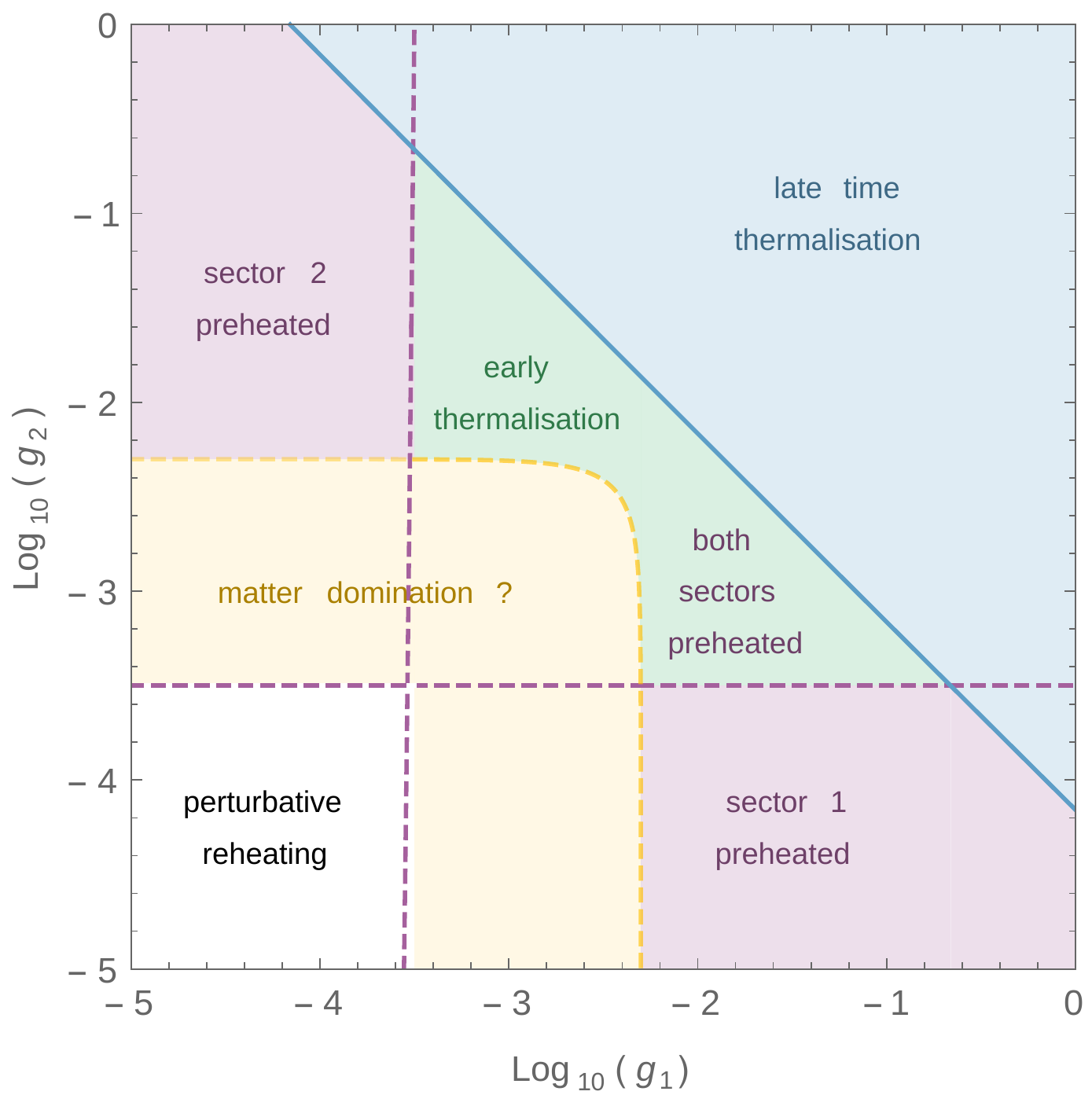}
\hspace{4mm}
\includegraphics[width=0.46\textwidth]{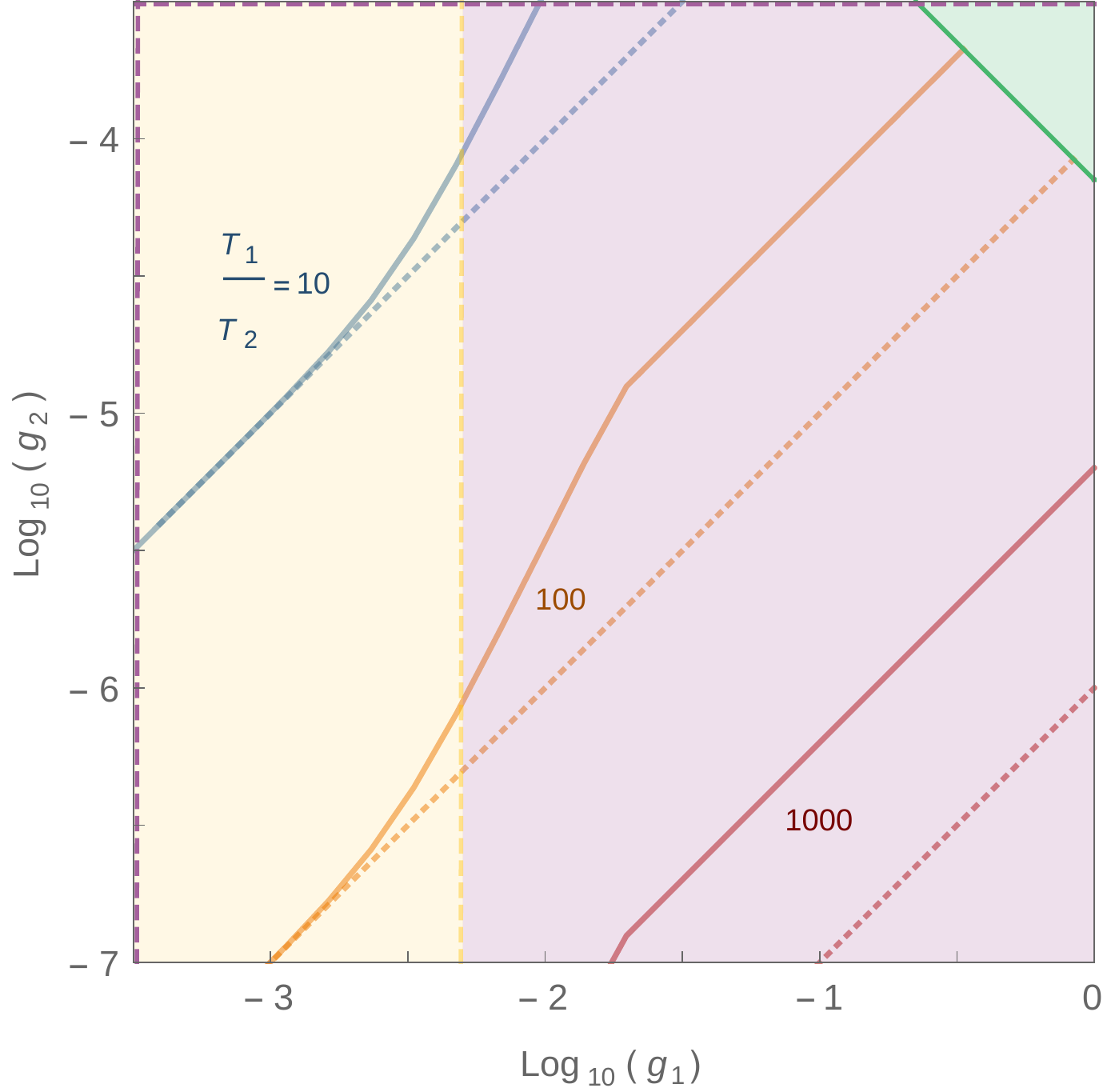}
\caption{{\bf \emph{Left:}}  Late time effects in a model with an inflaton $\phi$ of mass $10^{13}~\GeV$ and two  scalars  $\chi_{1,2}$ with mass $m_{\chi}=10^6~\GeV$, coupled to the inflaton by $\sum_i \frac{1}{2} g_i^2 \phi^2 \chi_i^2 + \frac{1}{2} g_i m_{\phi} \phi \chi_i^2$. Models to the right or above the purple dashed lines have at least one sector efficiently preheated. Inflaton mediated thermalisation can happen at late times, when the temperature is $\simeq m_{\chi}$. For small $g_i$ reheating is purely perturbative, or there is a period of matter domination leading to a temperature ratio close to perturbative expectation. Early time thermalisation is possible if both sectors are preheated. If only one sector is preheated, and is not thermalised with the other, Bose enhancement of the late time inflaton decays can have a significant effect on the relative reheat temperatures. {\bf \emph{ Right:}} Plot of the final temperature ratio between sectors in the part of parameter space where only one 
sector is preheated (solid lines). The model is the same as the left panel, and we assume that the preheated sector internally thermalises at $t\sim 10^4/m_{\phi}$, and that inflaton decays are perturbative after this and not efficient before.  The Bose enhancement of inflaton decay can cause significant deviations from the perturbative prediction (dotted lines).
\label{fig:therm}}
\end{figure*}

As an example we focus on the quadratic inflation model, Eq.~\eqref{eq:trilineaL}, with SUSY inspired trilinear and quartic couplings to the other fields. Throughout we take $m_{\phi}=10^{13}~\GeV$, and $\chi_1$ and $\chi_2$ to have the same mass $m_{\chi}$, detailed study of other models would also be interesting, but is left to future work. A summary of the results for this model, with $m_{\chi}=10^6~\GeV$ is shown in Fig.~\ref{fig:therm}. 

\subsection{Late time thermalisation}

For sufficiently large inflaton trilinear couplings, separate sectors will thermalise with each other at temperatures below the perturbative reheat temperature, mediated by inflaton scattering. If this is the case, at late times the two sectors will have comparable energy densities, regardless of earlier effects. The details of this computation have been studied carefully in \cite{Adshead:2016xxj}, where explicit formula for the scattering rates are given. The relative energy transfer rate from sector 1 to sector 2 via inflaton scattering is defined as
\beq \label{eq:scale}
\Gamma^{1\rightarrow2}\left(T_1,T_2\right) = \frac{n\left(T_1\right)^2 \left<\sigma v E \right>\left(T_1,T_2\right)}{\rho_1}~,
\eeq
where $\rho_1$ is the energy density in sector 1, which is taken to be at higher temperature than sector 2. If $\Gamma^{1\rightarrow2}\left(T_1,T_2\right) $ exceeds the Hubble parameter at any time, the two sectors will thermalise. 

In the model Eq.~\eqref{eq:trilineaL}, the trilinear couplings are dimensionful, so thermalisation will happen most easily at the lowest temperatures satisfying $T \gtrsim m_{\chi}$. Provided $m_{\chi}$ is not close to the predicted perturbative reheat temperature, both sectors will be close to a thermal distribution at this time, and at temperatures below the inflaton mass an approximate scaling
\beq \label{eq:Gapprox}
\frac{\Gamma^{1\rightarrow 2}\left(T_1,0\right)}{H} \sim \frac{g_1^2 g_2^2 M_{\rm Pl}}{T_1}~,
\eeq
is expected \cite{Adshead:2016xxj}. This dependence is only a rough estimate, and we obtain numerical results  by computing the relevant scattering processes. In models with other types of couplings, for example an inflaton coupled to a fermion, thermalisation will be dominated by high temperatures.

In Fig.~\ref{fig:therm} we show the parts of parameter space thermalised when the decay products both have a mass of $10^6~\GeV$ (and the inflaton mass is $m_\phi\sim10^{13}~\GeV$).\footnote{If one of the sectors has typical mass scale much higher than the other, thermalisation is expected to dominantly occur when the hotter sector has a temperature close to the heavier mass scale (assuming scalars). However, a full computation is required for reliable results.} Because the dominant energy transfer happens at low temperatures, Bose enhancement of the cross section is not important. Consequently, the results are effectively the same if the constraint $\Gamma^{1\rightarrow2}\left(T_1,T_2\right) > H$ is imposed with $T_1 = T_2$ directly, rather than starting at early times with the post-reheating value of $T_2$, and tracking its evolution with time. For the model we consider, a large part of the parameter space with at least one sector preheated is thermalised. The effect of changing the decay product masses in 
a full computation is well reproduced by the scaling Eq.~(\ref{eq:Gapprox}) after setting $T_1 = m_{\chi} $ (provided the reheat temperature is significantly below the inflaton mass), and the minimum value of the product of the coupling constants for thermalisation (with $m_{\phi}=10^{13}~\GeV$ still fixed) is approximately
\beq
g_{1} g_2 \gtrsim 2\times10^{-7}  \left(\frac{m_{\chi}}{\GeV}\right)^{1/2} ~.
\eeq
There may be interesting phenomenological possibilities if a sector which starts off relatively cold, but is warmed up by thermalisation at late times. In the model with $m_{\chi} = 10^6~\GeV$, this only happens for the small part of parameter space in which only one sector is preheated, but thermalisation is efficient. However, it is possible over large parts of parameter space for smaller decay product masses.

\subsection{Matter domination}

If late time thermalisation does not occur, a possible period of matter domination will be the most important effect for the final relative temperatures of the two sectors. This happens if, after preheating, a significant population of non-relativistic inflaton states forms, and exists for an extended time before decaying perturbatively \cite{Giudice:1999fb,Podolsky:2005bw}. As a result, the energy transferred to the $\chi_i$ sectors is redshifted away, and the late time temperatures are just set by the perturbative prediction. In particular, any mass the $\chi_i$ states have from an expectation value at the end of preheating is expected to decrease as $\sim 1/a(t)$ where $a\left(t\right)$ is the scale factor of the universe, and redshifting alone will not cause the $\chi_i$ states to become more non-relativistic if their bare mass is small. As discussed in Section \ref{sec:3}, at the end of preheating the inflaton mass could also dominantly come from expectation values of thermal contributions if rescattering is important, but its bare mass is assumed larger than that of the $\chi_i$, so finite momentum quanta will become non-relativistic faster. Also, if interactions are not efficient at depleting the inflaton zero mode, this will immediately act as a matter component of the energy density.

The trilinear inflaton couplings in the model we study, Eq.~\eqref{eq:trilineaL}, are relatively large (for example compared to those in a model with couplings $\sim g_i^2 m_{\phi} \phi \chi_i^2$). This has two important effects. First it changes the perturbative inflaton decay rate, and the part of parameter space that is thermalised at late time, relative to the region that is preheated. The second, more subtle, effect is that numerical studies indicate that a large trilinear coupling causes the inflaton states after preheating to be far more relativistic than if the trilinear was absent \cite{Dufaux:2006ee}. 

In the case of a small trilinear, $\sim g^2$, lattice simulations show that from the time when preheating ends at $t \sim 100/m_{\phi}$ until times of $1000/m_{\phi}$ the fraction of $\chi$ states that are relativistic increases steadily to $\simeq 1/3$. Meanwhile, $\simeq 1/5$ of $\phi$ states are relativistic at $t=200/m_{\phi}$, and this subsequently drops to $\simeq 1/20$ at $t= 1000/m_{\phi}$ \cite{Podolsky:2005bw,Dufaux:2006ee}. At these times the equation of state parameter $w= p/\epsilon$ (where $p$ is the pressure and $\epsilon$ the energy density), is $w\simeq 0.15$ and is decreasing towards the matter dominated value. More recent simulations run for longer, show that in such a model the fraction of energy in $\chi$ states  drops approximately as $\sim 1/a\left(t\right)$ at late times. This shows that matter domination is reached relatively quickly, and barring extremely large values of $g_i$ at least some period of matter domination is inevitable. This is compatible with recent studies, which show that the matter dominated equation of state is quickly reached in a quadratic inflaton potential model without large interactions with decay products \cite{Lozanov:2016hid}. The dilution of the energy density in the $\chi_i$ produced by preheating is given by the $\left(t_{m}/t\right)^{2/3}$, where $t_m$ is the time at which matter domination sets in, estimated as $t_{m} \sim 250/m_{\phi}$.

In contrast, models with large trilinear couplings $\sim g$ have different behaviour. Up to the times of the longest simulations carried out so far, $t=1000/m_{\phi}$, the fraction of both inflaton and $\chi$ quanta that are relativistic are very similar, with both increasing to $\simeq 1/2$ steadily over the simulations \cite{Dufaux:2006ee}.\footnote{An even larger fraction are relativistic if the quartic is absent.} Meanwhile the equation of state is constant at $w\sim 0.25$, and there is a relatively fast flow of the energy and number density spectrum towards UV modes, and kinetic equilibrium. However, the distribution of modes is still fairly IR localised compared to a thermalised system, which would have temperature $\simeq 10^{14}~\GeV$. At these late times, the energy is still approximately equally distributed between the inflaton and an efficiently preheated sector, and therefore matter domination has not set in. We stress that by focusing on a model with large trilinears, we are choosing an option for which matter domination will have a relatively small effect, and in other cases it will be important over larger parts of parameter space.
 
Considering the model Eq.~\eqref{eq:trilineaL}, an estimate for the parameter range for which matter domination is important can be made by assuming that both the inflaton and $\chi$ states continue moving towards a thermal distribution (numerical simulations would be useful to verify if this is accurate). Consequently, matter domination will occur if the timescale for perturbative inflaton decays exceeds the time taken for the energy in the inflaton sector to redshift to an effective temperature $\lesssim m_{\phi}$, which happens at a time $\simeq 10^5/m_{\phi}$ (based on an equation of state close to the radiation dominated value).  We make the simple assumptions that perturbative inflaton decay rate (including a possible Bose enhancement) is accurate, and that any significant expectation value induced and thermal masses are negligible, which is plausible given that this is happening at relatively late times. Energy could also be extracted from the inflaton by scattering processes, which would need to be included in a complete analysis, along with a study of the quasiparticle structure of the plasma \cite{Drewes:2013iaa}.

Taking the potential of Eq.~(\ref{eq:trilineaL}), and assuming the $\chi$ are thermalised, the inflaton decay rate is
\beq \label{eq:ind}
\Gamma_{\phi\rightarrow \chi_i \chi_i} \left(T_i\right)  = 
\frac{g_i^2m_{\phi}}{32 \pi} \sqrt{1-\frac{4 m_{\chi}^2}{m_{\phi}^2}} \left(1+ \frac{2}{e^{m_{\phi}/\left(2T_i\right)}-1} \right) ~,
\eeq
where the factor in brackets is the Bose enhancement from the $\chi$ number density \cite{Adshead:2016xxj}. This expression is obtained from the decoherence rate of a single inflaton mode \cite{Koksma:2011dy}, assuming that its spatial momentum is small \cite{Drewes:2010pf} (which is a reasonable approximation since the majority of the quanta produced by rescattering have relatively small momentum). However, it does not include effects such as hydrodynamic slowdown, discussed in  \cite{Berges:2008sr,Berges:2016nru}, which could lead to significant corrections, but are beyond the scope of our present analysis. As the temperature of the hottest $\chi_i$ sector drops after preheating, there will be a time when $\Gamma_{\phi\rightarrow \chi_i \chi_i}\left(T_i\right) \simeq H\left(T_i\right)$  (the corresponding temperatures are significantly below the temperature immediately after preheating for typical values of $g_i$). For this to happen before matter domination, requires a coupling $g_i\gtrsim 0.005$.

An alternative plausible assumption is that the distribution of decay products could remain close to that at the end of preheating (at a time $t_p$), apart from the effect of the expansion of the universe, corresponding to no significant flow of modes to the UV. If this is the case, the enhancement relative to the decay rate when there is no Bose enhancement, is a factor of $f_{\chi}\left(\frac{m_{\phi}}{2} \frac{a\left(t\right)}{a\left(t_p\right)} \right)$, where $f_{\chi}\left(k\right)$ is the  occupation number distribution at a time $t_p$. From lattice results, in this scenario the enhancement can be large. For example at $t\simeq 1000/m_{\phi}$, the occupation number distribution is approximately
 \beq \label{eq:approxf}
 f_{\chi}\left(k_c\right) \simeq 10^{11} \left(\frac{m_{\phi}}{k_c}\right)^{-3.5} ~,
 \eeq
where $k_c$ is the comoving momentum of a mode defined relative to the beginning of preheating at a time $t_0 \simeq m_{\phi}$. In this case, decays are fastest compared to the Hubble parameter at early times due to the power dependence in Eq.~\eqref{eq:approxf}. If the inflaton modes are thermalised, then provided decays are possible by a time $10^4/m_{\phi}$, matter domination will not occur if $g_i \gtrsim 0.001$. In contrast, if there is no further flow of inflaton modes towards the UV, matter domination will happen earlier, within a time corresponding to $a\left(t\right)/a\left(1000/m_{\phi}\right) \sim 2$. Provided inflaton decays are not blocked, they could still be fast at these early time, however more complex dynamics, not well reproduced by perturbative calculations, might still lead to a period of matter domination.

Another possibility is that the inflaton has a quartic potential in the part of field space where reheating happens, although a pure quartic potential is ruled out by observations (it may be possible to improve agreement with non-minimal couplings to gravity \cite{Leontaris:2016jty}). In this case, matter domination does not occur because the inflaton sector energy density evolves as radiation. This is supported by \cite{Figueroa:2016wxr} who carry out a numerical simulation to relatively late times, finding an equation of state close to radiation and no relative decrease in the fraction of energy in inflaton decay products. Therefore, in models with a quartic potential, the final ratio of the temperatures of the sectors is determined by a combination of the energy distribution at early times, and the perturbative expectation from the order 1 fraction of energy remaining in the inflaton, if thermalisation does not take place. Recent studies \cite{Lozanov:2016hid} confirm this behaviour for any model in which 
the inflaton potential has form $ \phi^{n}$ with $n\gtrsim 2$.

\subsection{Early time thermalisation} \label{ettherm}

Another potentially important effect is inflaton mediated thermalisation at early times, enhanced by the high number densities immediately after preheating. Off-shell inflaton scattering in not included in the simulations of Section \ref{sec:3}, but can be calculated directly using the distributions from numerical simulations, assuming perturbative calculations of the scattering rate are reasonably accurate (thermalisation by production of on-shell inflaton states will be seen in the numerics if it happens on short enough timescales). The computation is very similar to \cite{Adshead:2016xxj}, with the difference that the energy distributions involved are very far from thermal, so that more integrals must be evaluated numerically. As before we compare the rate $\Gamma^{1\rightarrow2} \left(T_1,T_2\right)$ to the Hubble parameter (the rate at which the decay product's number distribution is altered by internal thermalisation could also give an important timescale, but this is unknown). If matter domination does not subsequently occur, early thermalisation will lead to similar final temperatures in the two sectors.

 Since Bose enhancement is potentially large, we calculate the rate of energy transfer using occupation number distributions from the numerics \cite{Dufaux:2006ee}, rather than assuming the two sectors already have equal energy density. A deficiency of this is that the number distributions are only giving in \cite{Dufaux:2006ee} for a model $g_i= 10^{-3}$, however many properties of preheating are not highly sensitive to the actual values of couplings provided they are above the threshold for efficient preheating. While we use the full distributions in our computations, as an indication the occupation numbers is approximately given by Eq.\eqref{eq:approxf} at $t \simeq 1000/m_{\phi}$. We again neglect any expectation value sourced masses, and there remains the possibility that effects due to the high occupation number of $\chi$ states could lead to large deviations, so our results should be taken as indicative.

In parts of parameter space where one sector is preheated, the enhancement from the relatively high number densities is not enough to overcome the much larger value of the Hubble parameter at these times. As a result, in this scenario, early time thermalisation does not happen apart from for very large couplings, far inside the region where late time thermalisation is efficient. However, if both sectors are preheated, there in a potentially enormous Bose enhancement, even though the normal cancellation in the rate equation means it is only by one factor of the decay product occupation number not two \cite{Adshead:2016xxj}. Assuming the dynamics of the system are such that the thermalisation rate is well approximated by a perturbative calculation at $t\sim 100/m_{\phi}$, this enhancement is as large as $\simeq 10^6$ for modes with physical momentum around $k\simeq m_{\phi}/20$ at this time (that is, modes with momentum around $m_{\phi}$ at the beginning of preheating). This is large enough to cause thermalisation over all of the parameter space where both sectors are preheated efficiently.

Meanwhile, if thermalisation is not possible until a later time, for example due to large expectation value sourced masses, it will be suppressed. This is because, although the Hubble parameter decreases, the number density of modes with momentum not far below the mass of the inflaton decreases fast due to redshifting as well. For example, assuming Eq.~\eqref{eq:approxf} is valid over a range of times, the occupation number of modes with physical momentum around $m_{\phi}$ drops as $a\left(t\right)^{-3}$. Supposing that the dynamics are such that thermalisation can first happen at a time $t\simeq 1000/m_{\phi}$, and both sectors get an order 1 fraction of the energy density, it occurs provided $g_1 g_2 \gtrsim 10^{-3}$. This is inside the late time thermalisation region, for $m_{\chi}= 10^6~\GeV$.

As an aside, we note that in the case of a preheated bosonic sector potentially thermalising with a non-preheated fermionic sector, a computation shows that the thermalisation rate at the end of preheating is enhanced compared to a thermal distribution with the same energy density. This leads to an especially large change in the thermalised parameter space, as in such models thermalisation typically requires temperatures $\gtrsim m_{\phi}$ \cite{Adshead:2016xxj}. Consequently, preheated thermalisation is stronger than thermalisation with same energy density, which itself is much stronger than thermalisation at the (relatively much lower) perturbative reheat temperature.

\subsection{Bose enhanced perturbative reheating}

Finally if none of these effects matter, the distribution of energy at the end of preheating can have an impact on the final ratio of reheat temperatures, beyond facilitating thermalisation. One possibility is that preheating could be the dominant source of energy for a relatively weakly coupled sector if perturbative inflaton decays are dominantly to another sector. This can happen because, provided its inflaton coupling is above a threshold, a sector typically gets an order 1 fraction of the total energy during preheating, even in the presence of a more strongly coupled sector. Alternatively,   a sector could be preheated, but not have significant trilinear couplings to the inflaton, preventing perturbative inflaton decays (or the inflaton could dominantly decay to a non-preheated fermionic sector).

Another possibility is that this high number density after preheating could alter the relative rate of perturbative inflaton decays to the two sectors.\footnote{Thermal and plasma dynamics can alter the final temperature ratio from $T_1/T_2 \sim g_1^{1/2}/g_2^{1/2}$ even in the absence of preheating, however the effect is smaller.}   The energy in the inflaton could also be distributed differently from the case of pure perturbative reheating, since after preheating a substantial proportion is in higher momentum modes rather than the zero mode, which could change the dynamics although we do not investigate this. Even in perturbative reheating scenarios, the dynamics can be modified significantly by thermal and plasma effects \cite{McDonald:1999hd,Davidson:2000er,Kolb:2003ke,Yokoyama:2004pf,Yokoyama:2005dv,Drewes:2010pf,Kurkela:2011ti,Mukaida:2012bz,Mukaida:2012qn,Mazumdar:2013gya,Drewes:2013iaa,Harigaya:2013vwa,Harigaya:2014waa,Drewes:2014pfa,Ho:2015jva,Lerner:2015uca,Koksma:2011dy}. Thermal masses produced as a sector heats up could modify or block energy transport \cite{Kolb:2003ke}, and energy transfer by scattering can be important, as can the quasiparticle and collective excitation dynamics  \cite{Drewes:2013iaa}. Quantifying these late time effects in a system with a complicated energy distributions left over from preheating is extremely challenging, even in the single sector case. A dedicated and computationally expensive numerical study is probably required, and this may not be within reach of current simulation power. 

Rather than attempting a full analysis of preheated models with multiple sectors, we instead focus on the simple possibility that a Bose enhancement of the inflaton zero mode decays can change the relative reheat temperatures of two sectors. While not likely to be representative of the full dynamics of a model, our calculation shows that the ratio of final temperatures can be significantly different to the simple perturbative expectation $ \sqrt{g_1 /g_2}$. In particular, we consider models in which only one sector is preheated, and therefore has  a high number density.   Unfortunately, even only considering the Bose enhancement, this depends on the unknown late time number distributions. However, it can be calculated if we assume that the states $\chi_i$ reach a thermal distribution fairly quickly. Whether this is accurate will depend on the other unspecified interactions in a sector.

Assuming internal thermalisation, the temperature of the preheated sector will drop until the inflaton decay rate is equal to the Hubble parameter, at a temperature $T_d$. The final temperature ratio is approximately given by the relative inflaton decay rates at this time
\beq \label{eq:boseratio}
\frac{T_1}{T_2} \sim \left(\frac{\Gamma_{\phi\rightarrow \chi_1 \chi_1}\left(T_d\right)}{\Gamma_{\phi\rightarrow \chi_2 \chi_2}\left(0\right)} \right) ^{1/4} ~.
\eeq

In Fig.~\ref{fig:therm} (right), we plot the  final ratio of the sectors temperatures over the part of parameter space in Fig.~\ref{fig:therm} (left) for which preheating happens to only one sector. Since sector 1 will not internally thermalise instantly after preheating, results are shown assuming this happens over a timescale $10^4/m_{\phi}$. We further assume that $\chi_1$ has a large mass from an expectation value such that the inflaton does not decay to it before this time, and that by this time the inflaton energy is in modes that are fairly close to non-relativistic so the Bose enhancement is $\simeq f_{\chi 1} \left(m_{\phi}/2\right)$. The simple perturbative prediction $\simeq \sqrt{g_1/g_2}$  for the relative temperatures of the two sectors is also shown for comparison.

For a particular point of parameter space, the ratio of temperatures can differ from $\sqrt{g_1/g_2}$ by a factor of $\sim 5$ (that is, up to $\sim 1000$ in the energy). The effect is most significant when the coupling $g_1$ is largest, and therefore the decay time is early when the first sector has a high temperature. We emphasise however that this is assuming the perturbative inflaton decay rate Eq.~\eqref{eq:ind} is valid, and the high occupation density of inflaton modes may lead to complex modifications not capture by Eq.~\eqref{eq:boseratio}. Rather than claiming precise results, our point instead is simply that this is an effect that can cause large deviations from the expected temperature ratio.

If the comoving distribution of $\chi$ remains close to its form at the end of preheating, the relative change will depend on the occupation number at the time of decay. In this scenario, decays are more efficient at early times, due to the redshifting of the momentum distribution. The relative temperatures are fixed by the timescale over which decays from the inflaton are possible, for example due to a large expectation value source mass for the $\chi$ decreasing. This depends on the details of the complex dynamics, and could lead to larger temperature ratios than assuming thermalisation.

\section{Non-Oscillatory models and large temperature asymmetries}
\label{sec:6}

In the models studied so far, large temperature ratios still require dramatic differences in sectors' couplings to the inflaton, although the ratio can be significantly different from the perturbative prediction. This is simply because backreaction and rescattering ends preheating when there is an order 1 fraction of the total energy left in inflaton modes. However, in Non-Oscillatory inflation models \cite{Spokoiny:1993kt,Joyce:1997fc,Peebles:1998qn,Felder:1999pv} the inflaton potential does not have a minimum, and instead is constant at large field values. As a result, the inflaton does not oscillate and perturbative decays are absent. Early Non-Oscillatory models had inefficient reheating, leading to unrealistic cosmologies, however the Instant Preheating mechanism can lead to sufficient reheating \cite{Felder:1998vq,Felder:1999pv}. In this scenario a state $\chi$, with mass due to the inflaton expectation value, is resonantly produced. As the inflaton moves to the larger field values, the mass of $\chi$ increases, extracting more energy from the inflaton. If $\chi$ decays to other lighter states while heavy, sufficient energy can be transferred for successful reheating.
 
 In this section we show that such models can lead to very large temperature differences from order 1 differences in the coupling constants. Analogously to Fig.~\ref{fig:ph}, one sector could be above the threshold of resonance and reheated efficiently, whereas another similar sector just below threshold will have exponentially suppressed energy transfer. Of course, we do not claim this is the only, or the simplest, way to get large temperature asymmetries. For example perturbative inflaton decay to a sector could be forbidden if the inflaton is not coupled to any lighter states in that sector  (although this constrains model building) or if it has no trilinear interactions \cite{Allahverdi:2002nb}. In these scenarios the possibility of a sector being strongly preheated but not reheated might allow for interesting phenomenology \cite{Chung:1998zb}.
 
We study a simple example model in which the inflaton couples to two scalars $\chi_i$ in different sectors. These are themselves coupled to fermions $\psi_i$ in their own sectors, so that the energy transfer is  $\phi \rightarrow \chi_i \rightarrow \psi_i$. The $\chi_i$ must only couple strongly to the corresponding $\psi_i$. The advantage, compared to having the inflaton itself coupled strongly to only one sector, is that $\chi_i$ can easily be charged, for example under an unbroken (or weakly broken) gauge symmetry. This makes model building strong decays of $\chi_i$ to only one sector straightforward. In contrast, attempting to make the inflaton part of the visible sector, in such a way that prevents it decaying to a generic hidden sector is more difficult \cite{Wang:2013qti}.

For definiteness we assume the simple Non-Oscillatory model studied in \cite{Felder:1998vq,Felder:1999pv} with a potential $V(\phi)=V_0+V_{\rm int}$ where
\beq
V_0(\phi)=\left\{\begin{array}{cc}
\frac{m^2}{2}|\phi|^2 & ~~~\phi<0\\[5pt]
0 & ~~~\phi>0\\
\end{array}
\right.~,
\eeq 
and
\beq
V_{\rm int}(\phi)= g_i^2\phi^2|\chi_i|^2 + m_i^2 |\chi_i|^2~.
\eeq
The inflaton is assumed to start at large negative field values, and moves towards $\phi=0$. At this point, the dynamics can become non-adiabatic allowing $\chi_i$ to be efficiently produced. Analogously to normal preheating, the number density generated is
\beq \label{eq:nonc}
n_{\chi i}=\frac{(g_i \dot \phi_0)^{\nicefrac{3}{2}}}{8\pi^3}\exp\left(-\frac{\pi m_i^2}{g_i|\dot \phi_0|}\right)~,
\eeq
where $\dot \phi_0$ is the inflaton velocity near the origin. Following \cite{Felder:1998vq}, in a model where inflation ends with $\phi \simeq -0.3 M_{\rm Pl}$, the inflaton speed at the origin is 
\beq
  |\dot\phi_0|\sim 10^{-7} M_{\rm Pl}^2~.
  \label{dot}
  \eeq  
From Eq.~\eqref{eq:nonc}, the production of $\chi_i$ states is exponential suppressed if
  \beq
  m_i> \sqrt{g_i|\dot \phi_0|/\pi}~,
  \label{cond}
  \eeq 
where the exponential suppression is because the theory is close to evolving adiabatically if this condition is satisfied. For a coupling of $g_i\sim 1$ the threshold in this model is $m_i\sim 10^{15}~\GeV$. Since particle production only happens once, when $\phi$ passes through the origin, this result is not affected by the presence of a second sector. Subsequently, $\phi$ evolves to larger values. Although $n_{\chi i}$ is constant, the mass of $\chi_i$ grows $\sim g_i \phi(t)$, and the $\chi_i$ energy density increases
\beq
\rho_{\chi i}(t)=\frac{(g_i \dot \phi_0)^{\nicefrac{3}{2}}}{8\pi^3}\frac{g_i \phi(t)}{a(t)^3} \exp\left(-\frac{\pi m_i^2}{g_i|\dot \phi_0|}\right)~.
\label{instrho}
\eeq 
Meanwhile, energy remaining in the inflaton redshifts as $a(t)^{-6}$, because it is in the form of kinetic energy  \cite{Spokoiny:1993kt}, so does not dominate the universe at late times. 

The produced $\chi_i$ states affect the inflaton equation of motion, potentially making it move back towards $\phi=0$, which would cause $\rho_{\chi i}$ to decrease. We consider the simplest possibility, which is that the $\chi_{i}$ decay to other lighter states before this happens. For definiteness, if $\chi_i$ is coupled to significantly lighter fermions through an interaction $\kappa_i\chi_i\psi_i\bar{\psi}_i$, the decay rate is
\beq
\Gamma_{\chi_i\rightarrow\psi_i\overline{\psi}_i}\simeq\frac{\kappa_i^2g_i|\phi(t)|}{8\pi}~.
\eeq
It can be shown that provided 
\beq \label{eq:nocd}
\kappa_i \gtrsim 0.005 g^{3/4} ~,
\eeq
decays happen sufficiently fast that the inflaton does not change direction \cite{Felder:1998vq} (quintessence style models where the potential is not exactly zero for $\phi>0$ may be viable, and can relax this requirement).

The total relative energy transferred to each sector (evaluated at late times) is therefore given by
\beq \label{eq:noen}
\frac{\rho_1}{\rho_2} \simeq \left(\frac{g_1 \kappa_2}{g_2 \kappa_1}\right)^{5/3}  \exp\left(-\frac{\pi}{|\dot \phi_0|}\left(\frac{m_1^2}{g_1}-\frac{m_2^2}{g_2}\right)\right) ~,
\eeq
assuming matter domination between the decays of $\chi_1$ and $\chi_2$, and making the approximation that $\dot{\phi}$ is constant until both the $\chi_1$ and $\chi_2$ have decayed, for simplicity.\footnote{The speed of $\phi$ will actually decrease, and  $\phi = M_{\rm Pl}/\left(2\sqrt{3}\right) \log\left(t/t_0\right) $ with $t_0 = 5/\left(\sqrt{3\pi} m_{\phi}\right)$ due to the expansion of the universe \cite{Felder:1999pv}. For relatively large $g_i$ and $\kappa_i$, the effect of this will be small on the timescale of $\chi_i$ decays. For smaller coupling constants, it will be an order 1 difference to Eq.~\eqref{eq:noen}. However, the overall form of a relatively weakly varying function of the couplings multiplied by an exponential suppression below threshold remains.} It can be seen from Eq.~\eqref{eq:noen} that if both sectors are in a part of parameter space where the number density Eq.~\eqref{eq:nonc} is not exponentially suppressed, they will get comparable energy density, with differences only from powers of coupling constants. In contrast, if one of the sectors is in the suppressed region it is colder with the energy ratio an exponential function of its coupling to the inflaton. The transition between these regimes only requires order 1 changes in the parameters of the two models, and also does not need tuning onto particular points in parameter space.\footnote{The more strongly reheated sector could also be in the exponentially suppressed region, however in this case it must be fairly close to the boundary, so that sufficient energy is transferred to be reheated to a high enough temperature for acceptable phenomenology.}

Given that these models typically involve relatively large couplings, both between the inflaton and the $\chi_i$, and between the $\chi_i$ and $\psi_i$, there is the potential for sectors to thermalise through inflaton scattering. However, thermalisation is only possible at very early times before the $\chi_i$ decay, or when the typical energy of $\psi_i$ states is high enough to produce a population of $\chi_i$. Imposing that Eq.~\eqref{eq:nocd} is satisfied, a negligible fraction of the  energy will be transferred to the other sector on these timescales. In particular, the mass of $\chi_{1,2}$ is bigger than that of the inflaton, so any on-shell inflaton states produced cannot decay and transfer energy to these.  The fraction of energy transferred by off-shell inflaton scattering on the relevant timescales, defined analogously to Eq.~\eqref{eq:scale}, is negligible as well. This is because the number density Eq.~\eqref{eq:nonc} is relatively low reducing scattering events compared to decays, and also the matrix element for s-channel scattering via an off-shell inflaton is suppressed since the center of mass energy is very large relative to the inflaton mass, while t-channel scattering is suppressed by the low number density of $\chi_i$ in a sector for which reheating is exponentially small. Meanwhile, the inflaton continues moving to larger field values, so the masses of the $\chi_i$ increase. Once these have grown by a factor $\sim 2$, production of $\chi_i$ from scattering of $\psi_i$ is no longer possible. This also happens on a timescale $ \Gamma_{\chi_i\rightarrow\psi_i\overline{\psi}_i}^{-1}$ since the inflaton is moving at an approximately constant speed. Consequently, as long as the inflaton only has large couplings to the $\chi_i$,  inflaton thermalisation mediated does not occur at later times either.


\section{Discussion and phenomenological implications}
\label{sec:7}

For many beyond the SM scenarios the relative temperatures of two sectors at very late times is crucial. As an example we showed that for the DM relic abundance to be set purely by freeze-in, a temperature ratio of $\gtrsim 10$ is needed between the visible and dark sectors. Possible temperature asymmetries between sectors are also important for understanding cosmological bounds on new light states  (for example, bounds on $\Delta N_{\rm eff}$ or energy injection to the visible sector). Low temperature hidden sectors might contain exotic states that could be detectable with future experiments, but would be ruled out by cosmological bounds in the absence of a temperature asymmetry \cite{Davidson:2000hf,Foot:2014uba}, although thermalisation through the coupling to the Standard Model must also be considered \cite{Vogel:2013raa}.  Some formulations of the relaxion mechanism \cite{Graham:2015cka}, which could  solve the Electroweak hierarchy problem, also need large temperature differences between the visible and dark sectors  \cite{Hardy:2015laa}. 

In other models, the details of a thermal history of a sector, rather than its late time relative temperature, can have implications. Possibilities include restoration of a broken symmetry at high temperatures, which might only happen if a sector is preheated \cite{Kofman:1995fi}. In this case, an understanding of the effects of multiple sectors on preheating is important. Similarly, whether preheating occurs can affect the production of very heavy, or very weakly coupled states \cite{Giudice:2001ep}, and potentially baryogenesis \cite{Kolb:1996jt,GarciaBellido:1999sv}. In another direction, interesting phenomenology might come from the possibility that a sector is initially cold after reheating, but is heated up by inflaton mediated thermalisation at late times.


\subsection{Implications for dark matter}
\label{sec:5}

Preheating and related effects can have a significant impact on the DM relic abundance. This depends on the details of the particular DM candidate and production mechanism, and we simply consider a few simple scenarios, which by no means cover all the possibilities. 

As a example suppose the DM is in a hidden sector uncoupled to the visible sector (except for inflaton mediated interactions), and that there is a light hidden sector state that the DM can annihilate to, similarly to the models of Section~\ref{2b}. The DM relic abundance is then given by Eq.~\eqref{eq:fi2}, and is approximately proportional to the temperature ratio between the hidden and visible sectors at late times. This is determined by preheating, perturbative reheating, and possible inflaton mediated thermalisation. 
We also assume that the inflaton decays to scalars in the visible and hidden sectors through couplings of the form Eq.~\eqref{eq:trilineaL}, and fix the coupling to the visible sector to $g_1= 0.03$  so that the visible sector undergoes preheating.  

\begin{figure*}[t!]
 \begin{center}
\includegraphics[width=0.7\textwidth]{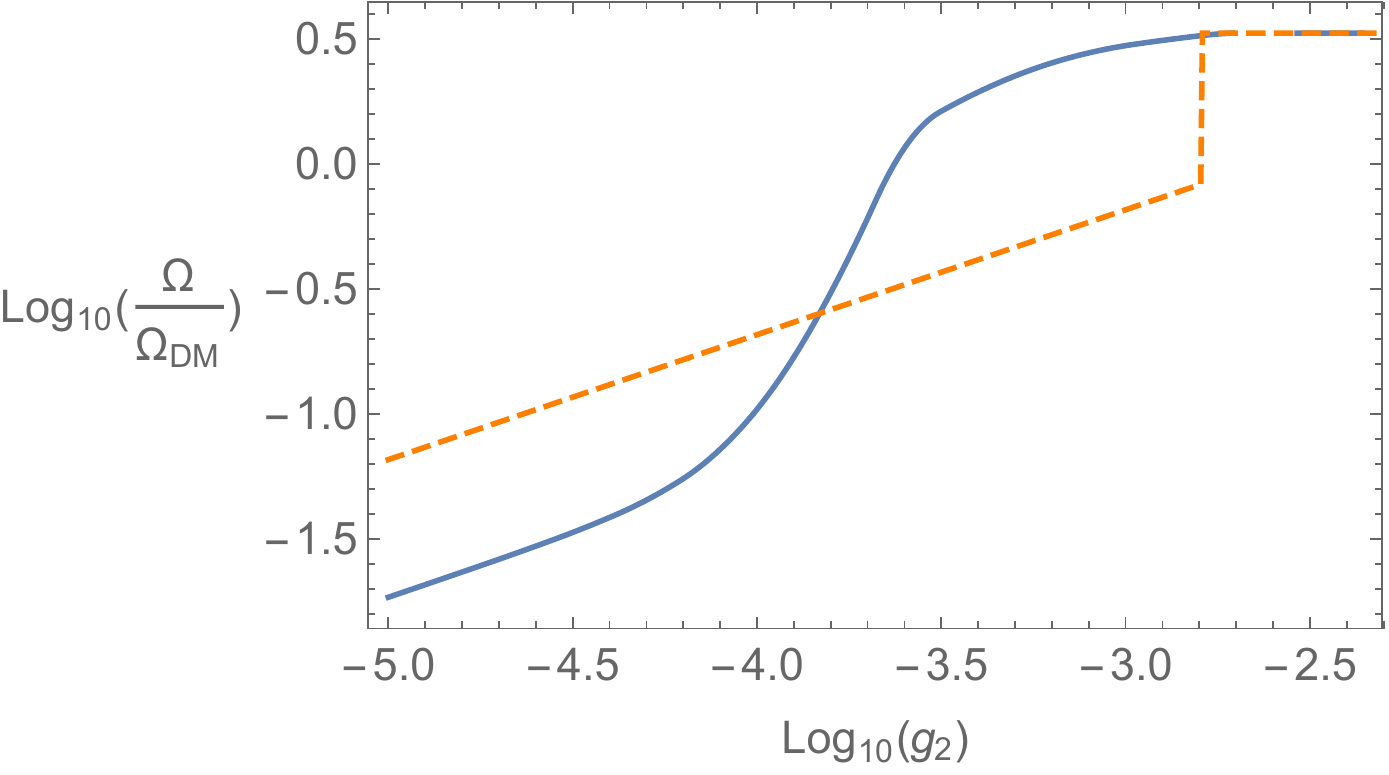}
 \end{center}
\caption{The DM relic abundance $\Omega$ compared to the observed value $\Omega_{\rm DM}$ in the example model described in the text, as the inflaton coupling to the hidden sector $g_2$ is varied. The solid line shows the result including the effects of preheating, while the dashed lines show the relic density that would be found if preheating was ignored and only perturbative inflaton decays considered. The inflaton coupling to the hidden sector is taken to have the form Eq.~\eqref{eq:trilineaL}, with a coupling constant $g_2$. Meanwhile the inflaton coupling to the visible sector has the same form, with a fixed coupling constant $g_1 = 0.03$, so that this sector is preheated. Additionally, we assume the DM can annihilate to light hidden sector states, with an  s-wave cross section $\left(\sigma v\right) = 0.01/ m_{DM}^2$, and that is has a mass of $3~\TeV$. For a given coupling to the inflaton, preheating can change the DM relic abundance by factors of order $10$. }
\label{fig:dm}
\vspace{2mm}
\end{figure*}

The DM relic abundance in such a model is plotted in Fig.~\ref{fig:dm} as a function of the coupling of the inflaton to the hidden sector $g_2$, for a DM candidate with mass $m_{DM} = 3~\TeV$ and an s-wave annihilation cross section $\left(\sigma v\right)=0.01/m_{DM}^2$. The results are shown with the effects of preheating included (solid line), and for comparison the relic density that would be obtained if the effects of preheating was neglected is also plotted (dashed line). The energy transfered during preheating is calculated using the {\tt LATTICEEASY} simulations of Section~\ref{sub3.5}. 
In this model Bose enhancement of perturbative inflaton decays to the visible sector, combined with the large trilinear couplings, means that perturbative reheating happens relatively fast, at times $t \sim 1000/m_{\phi}$. Consequently there is no period of matter domination. In computing the relic density we assume for definiteness that the energy in inflaton states redshifts as radiation between preheating and the time of perturbative decay, and the expansion of the universe is exactly radiation dominated at these times. However the results obtained are not very sensitive to these assumptions.

For very small values of $g_2$ less than approximately $10^{-4}$, the hidden sector is not preheated. In this case including preheating lowers the DM relic density relative to if its effects had been ignored. This is dominantly because the Bose enhancement increases the perturbative decay rate to the visible sector, decreasing the fraction of the energy transfered to the DM sector. The relative decrease in the hidden sector energy density is by a factor of approximately
\beq \label{eq:benh}
\frac{\Gamma_{\phi\rightarrow {\rm visible}}\left(0\right)}{\Gamma_{\phi\rightarrow {\rm visible}}\left(T_d\right)} \sim 0.01 ~,
\eeq
where $\Gamma_{\phi \rightarrow {\rm visible}}\left(T\right)$ is the inflaton decay rate at a temperature $T$, and $T_d$ is the visible sector temperature when perturbative decays dominantly occur, similarly to Eq.~\eqref{eq:boseratio}. The numerical result $0.01$ is dependent on the particular model, and the magnitude of the Bose enhancement, and can vary from $1 \div 0.001$ for $0.001 \lesssim g_1 \lesssim 1$ . Consequently the DM relic abundance is reduced by a factor of approximately $\sim 0.01^{1/4}$ in the particular model we consider (the actual decrease is slightly larger since preheating transfers an order 1 fraction of the inflaton energy to the visible sector, which is then unavailable for perturbative decays).

As $g_2$ is increased preheating to the hidden sector begins to occur and transfers a significant proportion of the inflaton energy, increasing the hidden sector temperature and the DM relic abundance. This corresponds to the sharp rise in Fig.~\ref{fig:dm}, and leads to a larger relic abundance than if preheating was ignored. Additionally, once the hidden sector gets an order 1 fraction of the inflaton energy early time inflaton mediated thermalisation becomes efficient due to the high number densities in the two sectors, as studied in Section~\ref{ettherm}. When this happens the visible and hidden sector have the same temperature, apart from a small effect due to the differing perturbative inflaton decay rates (a fraction of which occur after the two sectors are not longer in thermal contact). 

As a result, for $g_2 \gtrsim 0.003$ the hidden and visible sectors have approximately the same temperature, and any further increase in $
g_2$ has no effect on the DM relic abundance.  Meanwhile, if preheating was neglected the hidden sector temperature would still be significantly below that of the visible sector, leading to a suppressed relic density. Finally for $g_2 \gtrsim 0.001$ late time thermalisation is efficient, and the hidden and visible sectors have the same final temperatures even if preheating is neglected. This leads to a sharp increase in the DM relic density for the dashed curve for which preheating is ignored.

Preheating will also have a significant effect on the relic abundance in other dark matter scenarios. One simple case is if the dark matter sector has no number changing interactions, similarly to Section~\ref{2a}. If the dark matter sector is not preheated but the visible sector is, then the Bose enhancement of the perturbative decays to the visible sector results in production of fewer DM states. Analogously to Fig.~\ref{fig:ph} right and Eq.~\eqref{eq:benh},  Bose enhancement increases the inflaton decay rate to the visible sector by factors of up to $1 \div 1000$ for couplings to the visible sector in the range $g_1 \gtrsim 0.001$ (assuming trilinear couplings of the form Eq.~\eqref{eq:trilineaL}). In such models the DM relic abundance is simply proportional to the branching fraction of the inflaton to the hidden sector, and consequently is reduced by the same factor due to the Bose enhancement.

Another class of models in which preheating can have a dramatic effect on the relic abundance is if the DM is very heavy, known as ``wimpzillas''. One possibility is that superheavy DM is generated directly during preheating through its coupling to the inflaton \cite{Chung:1998zb,Giudice:1999fb,Greene:1997ge}. In this case, in the absence of preheating there will be no DM relic abundance in parts of parameter space where the DM is too heavy to be produced by perturbative inflaton decays (production from the thermal bath is also extremely suppressed for such large DM masses).

Alternatively,  DM could be very heavy and not be coupled to the inflaton, but instead produced from the thermal bath after preheating or reheating. In this case preheating still has a dramatic effect. For a DM mass above the maximum temperature the universe reaches after inflation $T_{\rm max}$, its relic density is parametrically suppressed by $\exp\left(-m_{\rm DM}/T_{\rm max} \right)$ \cite{Kuzmin:1997jua}. After preheating the effective temperature of the universe is very large, typically of order $10^{15}~\GeV$ if the inflaton mass is $\sim 10^{13}~\GeV$. This can lead to significant production of even very heavy DM at this time, although the exact relic abundance depends on the details of interactions and thermalisation, which determine when production of DM from the thermal bath becomes efficient. 

Meanwhile, in the absence of preheating the maximum temperature of the universe is parametrically $T_{\rm max} \sim H_I^{1/4} M_{\rm Pl}^{1/4} T_{RH}^{1/2} $ where $H_I$ is the Hubble parameter during inflation and $T_{RH}$ is the perturbative reheat temperature \cite{Giudice:2000ex} (this temperature is reached before reheating completes). For a quadratic inflaton potential with energy density during inflation of $m_{\phi}^2 M_{\rm Pl}^2$, and inflaton decay through a coupling to scalars $g m_{\phi}\phi \psi^2$, perturbative decays lead to a $T_{\rm max} \sim g^{1/2}/\left(32 \pi\right)^{1/4} m_{\phi}^{1/2} M_{\rm Pl}^{1/2}$. To obtain the correct DM relic abundance in such models the exponential suppression must typically satisfy $\exp\left(-m_{\rm DM}/T_{\rm max} \right) \ll 1$. As a result, including the effects of preheating leads to an exponentially large change in the relic density. Equivalently, the dark matter mass required to obtain the correct relic density is changed by a relative factor $T_{\rm PH}/ T_{\rm pert}$, where $T_{\rm PH}$ is the maximum effective temperature after preheating, and $T_{\rm pert}$ the maximum temperature including only perturbative decays. This shift can easily be several orders of magnitude depending on the inflaton couplings in a particular model.


\subsection{Summary}

In this paper we have studied preheating and reheating when the inflaton is coupled to two otherwise decoupled sectors, and related effects such as inflaton mediated thermalisation between sectors. In particular, we have attempted to track the energy in separate sectors during each stage of the evolution of the universe, highlighting potentially important physical processes. If two sectors are both preheated this can result in final temperatures much closer than the perturbative prediction, either directly or due to efficient thermalisation, provided there is not a long period of matter domination before perturbative reheating. Meanwhile, if only one sector is preheated, thermal effects such as Bose enhancement of perturbative inflaton decays can alter the ratio of final temperatures.

Our analysis has two clear shortcomings. First, for many of the effects of interest, the underlying physics is complex. To make progress, we have had to make strong simplifying assumptions, and these are unlikely to fully reproduce the true dynamics of the system. However, despite being unable to obtain precise results, we have pointed out possibly interesting processes, the role of which could be clarified with further work. Second, we have restricted ourselves to very simple toy models, which are unlikely to be representative of the visible sector, or hidden sectors in interesting beyond SM physics scenarios.  Although we have concentrated on scalars coupled to the inflaton, models coupled via fermions or gauge fields would also be worthwhile to explore \cite{Adshead:2015pva}. Further, we have considered only sectors that are almost identical copies of each other, differing only in their coupling to the inflaton. Even with these simplifications, mapping out the dynamics in the $g_1$ vs $g_2$ plane is 
challenging. Again, further work to consider more general scenarios would be worthwhile.

Finally, we have noted that very large temperature asymmetries are possible in particular models of inflation and reheating, without requiring the inflaton coupling to different sectors to be vastly different.  Large temperature differences could be interesting, for example in allowing heavy DM candidates that would have too large relic density in a warm sector, and there could be other model building possibilities.


\subsection*{Acknowledgements}

We are grateful to Joe Bramante and Robert Lasenby for useful comments, and Gary Felder for assistance with {\tt LATTICEEASY}.

\bibliographystyle{JHEP}
\bibliography{reference}

\end{document}